\documentclass[11pt]{article}
\usepackage{fixltx2e} 
\usepackage{cmap} 
\usepackage[T1]{fontenc}
\usepackage[utf8]{inputenc}
\usepackage{palatino}
\usepackage{multirow}
\usepackage{ifthen}
\usepackage{amsmath,amssymb}
\usepackage{longtable,ltcaption,array}
\newlength{\DUtablewidth} 

\makeatletter
\usepackage{fullpage}
\usepackage{moresize}
\usepackage{graphicx}
\renewcommand{\ttfamily}{\fontfamily\ttdefault\selectfont\scriptsize}

\makeatother
\makeatletter

\RequirePackage{color}

\definecolor{DUcolor0}{rgb}{1.00,0.00,0.00}

\expandafter\providecommand\csname DUrolec1\endcsname[1]{\textit{#1}}

\expandafter\providecommand\csname DUroles2\endcsname[1]{\textit{#1}}

\expandafter\providecommand\csname DUroles1\endcsname[1]{\textit{#1}}

\makeatother

\providecommand{\DUadmonition}[2][class-arg]{%
  \ifcsname DUadmonition#1\endcsname%
    \csname DUadmonition#1\endcsname{#2}%
  \else
    \begin{center}
      \fbox{\parbox{0.9\textwidth}{#2}}
    \end{center}
  \fi
}

\providecommand*{\DUfootnotemark}[3]{%
  \raisebox{1em}{\hypertarget{#1}{}}%
  \hyperlink{#2}{\textsuperscript{#3}}%
}
\providecommand{\DUfootnotetext}[4]{%
  \begingroup%
  \renewcommand{\thefootnote}{%
    \protect\raisebox{1em}{\protect\hypertarget{#1}{}}%
    \protect\hyperlink{#2}{#3}}%
  \footnotetext{#4}%
  \endgroup%
}

\providecommand*{\DUrole}[2]{%
  \ifcsname DUrole#1\endcsname%
    \csname DUrole#1\endcsname{#2}%
  \else
    \ifcsname docutilsrole#1\endcsname%
      \csname docutilsrole#1\endcsname{#2}%
    \else%
      #2%
    \fi%
  \fi%
}

\providecommand*{\DUtitle}[2][class-arg]{%
  \ifcsname DUtitle#1\endcsname%
    \csname DUtitle#1\endcsname{#2}%
  \else
    \smallskip\noindent\textbf{#2}\smallskip%
  \fi
}

\providecommand*{\DUtransition}[1][class-arg]{%
  \hspace*{\fill}\hrulefill\hspace*{\fill}
  \vskip 0.5\baselineskip
}

\ifthenelse{\isundefined{\hypersetup}}{
  \usepackage[colorlinks=true,linkcolor=blue,urlcolor=blue]{hyperref}
  \urlstyle{same} 
}{}
\hypersetup{
  pdftitle={Haskell for OCaml programmers},
  pdfauthor={Raphael ‘kena’ Poss}
}

\title{\phantomsection%
  Haskell for OCaml programmers%
  \label{haskell-for-ocaml-programmers}}
\author{Raphael ‘kena’ Poss}
\date{March 2014}

\begin{document}
\maketitle

This post mirrors \href{http://blog.ezyang.com/2010/10/ocaml-for-haskellers/}{OCaml for Haskellers} from Edward Z. Yang (2010).

\phantomsection\label{contents}
\pdfbookmark[1]{Contents}{contents}
\tableofcontents

\DUtransition

\DUadmonition[note]{
\DUtitle[note]{Note}

The latest version of this document can be found online at
\url{http://science.raphael.poss.name/haskell-for-ocaml-programmers.html}.
Alternate formats:
\href{http://science.raphael.poss.name/haskell-for-ocaml-programmers.txt}{Source},
\href{http://science.raphael.poss.name/haskell-for-ocaml-programmers.pdf}{PDF}.
}

\section{Prologue%
  \label{prologue}%
}

Why write a new post when a clever reader could simply “read \href{http://blog.ezyang.com/2010/10/ocaml-for-haskellers/}{Edward's
post} backwards”?

It's about the different audience, really.

My experience is that programmers well-versed in Haskell,
or who learn Haskell as first language, tend to have a strong
background in mathematics and logical reasoning. For this audience,
the abstract equivalences between Haskell and OCaml are trivial and do
not bear repeating. For them, a post like Edward's that mosty focuses
on the “limitations” of OCaml compared to Haskell, and provides
succinct high-level descriptions of OCaml's unique features, is
sufficient.

In contrast, an OCaml programmer is often someone who has learned OCaml by
iterative widening of their programming toolbox from other
languages, perhaps without particular sensitivity to the aesthetics of
abstraction. I once belonged to such an audience; I have taught
programming to it, too.

For practical OCaml programmers, I found that \textbf{it often works
better to approach Haskell from its operational perspective}, rather
than the formal/equational approach typically taken in Haskell
tutorials. The merits of Haskell's powerful abstraction compositions,
in this context, are a matter best left to self-discovery, later in
the learning process.

For example, Haskell tutorials often focus early on
non-strict evaluation and strong typing. This is an unfortunate
starting angle for a seasoned OCaml programmer who already knows
Haskell's type system very well (OCaml has mostly the same) and who is
likely to care immediately about a clear time and space cost model for
his/her code, an expectation broken at first sight by non-strict
evaluation.

\section{Why you should learn Haskell%
  \label{why-you-should-learn-haskell}%
}

There are two set of important features found in Haskell and not in OCaml:
\emph{killer features} and \emph{acquired tastes}.

Killer features are those that can immediately enable better productivity,
without incurring a feeling of loss from leaving OCaml's world:
\begin{itemize}

\item layout-based code structuring;

\setlength{\DUtablewidth}{\linewidth}
\begin{longtable*}[c]{p{0.470\DUtablewidth}p{0.470\DUtablewidth}}

{\ttfamily \raggedright \noindent
\DUrole{c}{(*~ML:~begin/end~needed~to~scope\\
~~~nested~pattern~matches~*)}~\\
\DUrole{k}{match}~\DUrole{n}{v1}~\DUrole{k}{with}~\\
\DUrole{o}{|}~\DUrole{mi}{0}~\DUrole{o}{->}~\DUrole{k}{begin}~\\
~~~~~~~~\DUrole{k}{match}~\DUrole{n}{v2}~\DUrole{k}{with}~\\
~~~~~~~~\DUrole{o}{|}~\DUrole{mi}{1}~\DUrole{o}{->}~\DUrole{s2}{\textquotedbl{}hello\textquotedbl{}}~\\
~~~~~~~~\DUrole{o}{|}~\DUrole{mi}{2}~\DUrole{o}{->}~\DUrole{s2}{\textquotedbl{}world\textquotedbl{}}~\\
~~~~~~~\DUrole{k}{end}~\\
\DUrole{o}{|}~\DUrole{mi}{3}~\DUrole{o}{->}~\DUrole{s2}{\textquotedbl{}yay\textquotedbl{}}
}
 & 
{\ttfamily \raggedright \noindent
\DUrole{cm}{\{-~HS:~indentation~disambiguates\\
~~~nested~pattern~matches~-\}}~\\
\DUrole{kr}{case}~\DUrole{n}{v1}~\DUrole{kr}{of}~\\
~\DUrole{mi}{0}~\DUrole{ow}{->}~\DUrole{kr}{case}~\DUrole{n}{v2}~\DUrole{kr}{of}~\\
~~~~~~~\DUrole{mi}{1}~\DUrole{ow}{->}~\DUrole{s}{\textquotedbl{}hello\textquotedbl{}}~\\
~~~~~~~\DUrole{mi}{2}~\DUrole{ow}{->}~\DUrole{s}{\textquotedbl{}world\\
~3~->~\textquotedbl{}}\DUrole{n}{yay}\DUrole{s}{\textquotedbl{}}
}
 \\
\end{longtable*}

\item declarations after use;
\begin{quote}{\ttfamily \raggedright \noindent
\DUrole{nf}{v}~\DUrole{ow}{=}~\DUrole{n}{sum}~\DUrole{mi}{20}~\\
~~\DUrole{kr}{where}~~\DUrole{cm}{\{-~no~equivalent~in~OCaml~-\}}~\\
~~~~\DUrole{n}{sum}~\DUrole{n}{n}~\DUrole{ow}{=}~\DUrole{n}{n}~\DUrole{o}{*}~\DUrole{p}{(}\DUrole{n}{n}~\DUrole{o}{-}~\DUrole{mi}{1}\DUrole{p}{)}~\DUrole{o}{/}~\DUrole{mi}{2}
}
\end{quote}

\item readable type interfaces directly next to declarations;

\setlength{\DUtablewidth}{\linewidth}
\begin{longtable*}[c]{p{0.470\DUtablewidth}p{0.470\DUtablewidth}}

{\ttfamily \raggedright \noindent
\DUrole{c}{(*~in~mod.mli~*)}~\\
\DUrole{k}{val}~\DUrole{n}{f}~\DUrole{o}{:}~\DUrole{kt}{int}~\DUrole{o}{->}~\DUrole{kt}{int}~\DUrole{o}{->}~\DUrole{kt}{int}~\\
\DUrole{c}{(*~in~mod.ml~*)}~\\
\DUrole{k}{let}~\DUrole{n}{f}~\DUrole{o}{=}~\DUrole{o}{(+)}
}
 & 
{\ttfamily \raggedright \noindent
\DUrole{cm}{\{-~in~Mod.hs~-\}}~\\
\DUrole{nf}{f}~\DUrole{ow}{::}~\DUrole{kt}{Int}~\DUrole{ow}{->}~\DUrole{kt}{Int}~\DUrole{ow}{->}~\DUrole{kt}{Int}~\\
\DUrole{nf}{f}~\DUrole{ow}{=}~\DUrole{p}{(}\DUrole{o}{+}\DUrole{p}{)}
}
 \\
\end{longtable*}

\item operator and function overloading;

\setlength{\DUtablewidth}{\linewidth}
\begin{longtable*}[c]{p{0.470\DUtablewidth}p{0.470\DUtablewidth}}

{\ttfamily \raggedright \noindent
\DUrole{c}{(*~no~ad-hoc~polymorphism,\\
~~~(+)~vs.~(+.)~*)}~\\
\DUrole{k}{begin}~\\
~~\DUrole{n}{print\_string}~\\
~~~~~\DUrole{o}{(}\DUrole{n}{string\_of\_int}~\DUrole{o}{(}\DUrole{mi}{3}~\DUrole{o}{+}~\DUrole{mi}{2}\DUrole{o}{));}~\\
~~\DUrole{n}{print\_string}~\\
~~~~~\DUrole{o}{(}\DUrole{n}{string\_of\_float}~\DUrole{o}{(}\DUrole{mi}{3}\DUrole{o}{.}\DUrole{mi}{0}~\DUrole{o}{+.}~\DUrole{mi}{0}\DUrole{o}{.}\DUrole{mi}{14}\DUrole{o}{));}~\\
\DUrole{k}{end}
}
 & 
{\ttfamily \raggedright \noindent
\DUrole{cm}{\{-~show~and~(+)~are~overloaded~-\}}~\\
~\\
\DUrole{kr}{do}~\\
~\DUrole{n}{putStr}~\\
~~~~~\DUrole{p}{(}\DUrole{n}{show}~\DUrole{p}{(}\DUrole{mi}{3}~\DUrole{o}{+}~\DUrole{mi}{2}\DUrole{p}{))}~\\
~\DUrole{n}{putStr}~\\
~~~~~\DUrole{p}{(}\DUrole{n}{show}~\DUrole{p}{(}\DUrole{mf}{3.0}~\DUrole{o}{+}~\DUrole{mf}{0.14}\DUrole{p}{))}
}
 \\
\end{longtable*}

\item more flexibility on operator and function names;

\setlength{\DUtablewidth}{\linewidth}
\begin{longtable*}[c]{p{0.470\DUtablewidth}p{0.470\DUtablewidth}}

{\ttfamily \raggedright \noindent
\DUrole{k}{let}~\DUrole{o}{(\textasciicircum{}.)}~\DUrole{n}{a}~\DUrole{n}{b}~\DUrole{o}{=}~\DUrole{n}{a}~\DUrole{o}{\textasciicircum{}}~\DUrole{s2}{\textquotedbl{}.\textquotedbl{}}~\DUrole{o}{\textasciicircum{}}~\DUrole{n}{b}~\\
\DUrole{k}{let}~\DUrole{n}{v}~\DUrole{o}{=}~\DUrole{s2}{\textquotedbl{}hello\textquotedbl{}}~\DUrole{o}{\textasciicircum{}.}~\DUrole{s2}{\textquotedbl{}world\textquotedbl{}}~\\
~\\
\DUrole{c}{(*~only~punctuation~can~serve~as\\
~~~infix~operators~*)}~\\
~\\
\DUrole{c}{(*~\textquotedbl{}.\textquotedbl{}~is~not~a~valid~prefix~*)}
}
 & 
{\ttfamily \raggedright \noindent
\DUrole{p}{(}\DUrole{o}{\textasciicircum{}.}\DUrole{p}{)}~\DUrole{n}{a}~\DUrole{n}{b}~\DUrole{ow}{=}~\DUrole{n}{a}~\DUrole{o}{++}~\DUrole{s}{\textquotedbl{}.\textquotedbl{}}~\DUrole{o}{++}~\DUrole{n}{b}~\\
\DUrole{nf}{v}~\DUrole{ow}{=}~\DUrole{s}{\textquotedbl{}hello\textquotedbl{}}~\DUrole{o}{\textasciicircum{}.}~\DUrole{s}{\textquotedbl{}world\textquotedbl{}}~\\
~\\
\DUrole{nf}{dot}~\DUrole{n}{a}~\DUrole{n}{b}~\DUrole{ow}{=}~\DUrole{n}{a}~\DUrole{o}{\textasciicircum{}.}~\DUrole{n}{b}~\\
\DUrole{nf}{v2}~\DUrole{ow}{=}~\DUrole{s}{\textquotedbl{}hello\textquotedbl{}}~\DUrole{p}{`}\DUrole{n}{dot}\DUrole{p}{`}~\DUrole{s}{\textquotedbl{}world\textquotedbl{}}~\\
~\\
\DUrole{p}{(}\DUrole{o}{.}\DUrole{p}{)}~\DUrole{n}{f}~\DUrole{n}{g}~\DUrole{n}{x}~\DUrole{ow}{=}~\DUrole{n}{f}~\DUrole{p}{(}\DUrole{n}{g}~\DUrole{n}{x}\DUrole{p}{)}
}
 \\
\end{longtable*}

\item configurable operator associativity and precedence.

\setlength{\DUtablewidth}{\linewidth}
\begin{longtable*}[c]{p{0.470\DUtablewidth}p{0.470\DUtablewidth}}

{\ttfamily \raggedright \noindent
\DUrole{k}{let}~\DUrole{o}{(/\textasciitilde{})}~\DUrole{n}{a}~\DUrole{n}{b}~\DUrole{o}{=}~\DUrole{n}{a}~\DUrole{o}{/}~\DUrole{n}{b}~\\
\DUrole{k}{let}~\DUrole{o}{(-\textasciitilde{})}~\DUrole{n}{a}~\DUrole{n}{b}~\DUrole{o}{=}~\DUrole{n}{a}~\DUrole{o}{-}~\DUrole{n}{b}~\\
~\\
\DUrole{c}{(*~all~operators~starting~with~the\\
~~~same~character~have~the~same\\
~~~associativity~and~precedence~*)}~\\
~\\
\DUrole{k}{let}~\DUrole{n}{v}~\DUrole{o}{=}~\DUrole{mi}{12}~\DUrole{o}{/\textasciitilde{}}~\DUrole{mi}{3}~\DUrole{o}{-\textasciitilde{}}~\DUrole{mi}{1}~\DUrole{o}{/\textasciitilde{}}~\DUrole{mi}{2}~\\
~~~\DUrole{c}{(*~=~(12~/\textasciitilde{}~3)~-\textasciitilde{}~(1~/\textasciitilde{}~2)~*)}~\\
~~~\DUrole{c}{(*~=~4~-\textasciitilde{}~0~*)}~\\
~~~\DUrole{c}{(*~=~4~*)}
}
 & 
{\ttfamily \raggedright \noindent
\DUrole{p}{(}\DUrole{o}{/\textasciitilde{}}\DUrole{p}{)}~\DUrole{n}{a}~\DUrole{n}{b}~\DUrole{ow}{=}~\DUrole{n}{a}~\DUrole{o}{/}~\DUrole{n}{b}~\\
\DUrole{p}{(}\DUrole{o}{-\textasciitilde{}}\DUrole{p}{)}~\DUrole{n}{a}~\DUrole{n}{b}~\DUrole{ow}{=}~\DUrole{n}{a}~\DUrole{o}{-}~\DUrole{n}{b}~\\
~\\
\DUrole{kr}{infixr}~\DUrole{mi}{7}~\DUrole{o}{-\textasciitilde{}}~~~\DUrole{cm}{\{-~higher~=~tighter~-\}}~\\
\DUrole{kr}{infixr}~\DUrole{mi}{6}~\DUrole{o}{/\textasciitilde{}}~~~\DUrole{cm}{\{-~r~for~right-assoc~-\}}~\\
~\\
\DUrole{nf}{v}~\DUrole{ow}{=}~\DUrole{mi}{12}~\DUrole{o}{/\textasciitilde{}}~\DUrole{mi}{3}~\DUrole{o}{-\textasciitilde{}}~\DUrole{mi}{1}~\DUrole{o}{/\textasciitilde{}}~\DUrole{mi}{2}~\\
~~~\DUrole{cm}{\{-~=~12~/\textasciitilde{}~((3~-\textasciitilde{}~1)~/\textasciitilde{}~2)~~-\}}~\\
~~~\DUrole{cm}{\{-~=~12~/\textasciitilde{}~(2~/\textasciitilde{}~2)~-\}}~\\
~~~\DUrole{cm}{\{-~=~12~/\textasciitilde{}~1~-\}}~\\
~~~\DUrole{cm}{\{-~=~12~-\}}
}
 \\
\end{longtable*}

\end{itemize}

“Acquired tastes” are Haskell features that a) can yield a significant
productivity improvement in the long run b) can yield terrible
performance and/or unreadable code when used naively and thus c) will
require patience and practice until you start feeling comfortable
using them:
\begin{itemize}

\item pure arrays;

\item function definitions by equations over infinite data structures;

\item type classes and custom Functor / Applicative / Monoid instances.

\end{itemize}

Some examples are given later below.

\DUtransition

\section{Straightforward equivalences%
  \label{straightforward-equivalences}%
}

Naming of types:
\begin{quote}{\ttfamily \raggedright \noindent
int~float~char~string~bool~userDefined~(*~OCaml~*)\\
Int~Float~Char~String~Bool~UserDefined~\{-~Haskell~-\}
}
\end{quote}

Operators:
\begin{quote}{\ttfamily \raggedright \noindent
=~~~<>~~\textasciicircum{}~~~@~~~@@~~+.~~/.~~*.~~+.~~(*~OCaml~*)\\
==~~/=~~++~~++~~\$~~~+~~~/~~~*~~~+~~~\{-~Haskell~-\}\\
~\\
land~lor~lxor~{[}la{]}sl~{[}la{]}sr~lnot~~~~~~~(*~OCaml~*)\\
.\&.~~.|.~xor~~shiftL~shiftR~complement~\{-~Haskell~-\}
}
\end{quote}

Functions:

\setlength{\DUtablewidth}{\linewidth}
\begin{longtable*}[c]{p{0.470\DUtablewidth}p{0.470\DUtablewidth}}

{\ttfamily \raggedright \noindent
\DUrole{k}{let}~\DUrole{n}{f}~\DUrole{n}{x}~\DUrole{n}{y}~\DUrole{o}{=}~\DUrole{n}{x}~\DUrole{o}{+}~\DUrole{n}{y}~\\
\DUrole{k}{let}~\DUrole{n}{g}~\DUrole{o}{=}~\DUrole{k}{fun}~\DUrole{n}{x}~\DUrole{n}{y}~\DUrole{o}{->}~\DUrole{n}{x}~\DUrole{o}{+}~\DUrole{n}{y}~\\
~\\
\DUrole{k}{let}~\DUrole{k}{rec}~\DUrole{n}{fact}~\DUrole{n}{n}~\DUrole{o}{=}~\\
~~~~\DUrole{k}{if}~\DUrole{n}{n}~\DUrole{o}{=}~\DUrole{mi}{1}~\DUrole{k}{then}~\DUrole{mi}{1}~\\
~~~~\DUrole{k}{else}~\DUrole{n}{n}~\DUrole{o}{*}~\DUrole{n}{fact}~\DUrole{o}{(}\DUrole{n}{n}\DUrole{o}{-}\DUrole{mi}{1}\DUrole{o}{)}~\\
~\\
\DUrole{k}{let}~\DUrole{k}{rec}~\DUrole{n}{fact'}~\DUrole{o}{=}~\DUrole{k}{function}~\\
~~\DUrole{o}{|}~\DUrole{mi}{1}~\DUrole{o}{->}~\DUrole{mi}{1}~\\
~~\DUrole{o}{|}~\DUrole{n}{n}~\DUrole{o}{->}~\DUrole{n}{n}~\DUrole{o}{*}~\DUrole{n}{fact'}~\DUrole{o}{(}\DUrole{n}{n}\DUrole{o}{-}\DUrole{mi}{1}\DUrole{o}{)}
}
 & 
{\ttfamily \raggedright \noindent
\DUrole{nf}{f}~\DUrole{n}{x}~\DUrole{n}{y}~\DUrole{ow}{=}~\DUrole{n}{x}~\DUrole{o}{+}~\DUrole{n}{y}~\\
\DUrole{nf}{g}~\DUrole{ow}{=}~\DUrole{nf}{\textbackslash{}}\DUrole{n}{x}~\DUrole{ow}{->}~\DUrole{nf}{\textbackslash{}}\DUrole{n}{y}~\DUrole{ow}{->}~\DUrole{n}{x}~\DUrole{o}{+}~\DUrole{n}{y}~\\
~\\
\DUrole{nf}{fact}~\DUrole{n}{n}~\DUrole{ow}{=}~\DUrole{cm}{\{-~no~\textquotedbl{}rec\textquotedbl{}~needed~-\}}~\\
~~~~\DUrole{kr}{if}~\DUrole{n}{n}~\DUrole{ow}{=}~\DUrole{mi}{1}~\DUrole{kr}{then}~\DUrole{mi}{1}~\\
~~~~\DUrole{kr}{else}~\DUrole{n}{n}~\DUrole{o}{*}~\DUrole{n}{fact}~\DUrole{p}{(}\DUrole{n}{n}\DUrole{o}{-}\DUrole{mi}{1}\DUrole{p}{)}~\\
~\\
\DUrole{cm}{\{-~equational:~order~matters~-\}}~\\
\DUrole{nf}{fact'}~\DUrole{mi}{1}~\DUrole{ow}{=}~\DUrole{mi}{1}~\\
\DUrole{nf}{fact'}~\DUrole{n}{n}~\DUrole{ow}{=}~\DUrole{n}{n}~\DUrole{o}{*}~\DUrole{n}{fact'}~\DUrole{p}{(}\DUrole{n}{n}\DUrole{o}{-}\DUrole{mi}{1}\DUrole{p}{)}
}
 \\
\end{longtable*}

Unit:

\setlength{\DUtablewidth}{\linewidth}
\begin{longtable*}[c]{p{0.470\DUtablewidth}p{0.470\DUtablewidth}}

{\ttfamily \raggedright \noindent
\DUrole{c}{(*~val~f~:~unit~->~int~*)}~\\
\DUrole{k}{let}~\DUrole{n}{f}~\DUrole{bp}{()}~\DUrole{o}{=}~\DUrole{mi}{3}
}
 & 
{\ttfamily \raggedright \noindent
\DUrole{nf}{f}~\DUrole{ow}{::}~\DUrole{nb}{()}~\DUrole{ow}{->}~\DUrole{kt}{Int}~\\
\DUrole{nf}{f}~\DUrole{nb}{()}~\DUrole{ow}{=}~\DUrole{mi}{3}
}
 \\
\end{longtable*}

Pattern match and guards:

\setlength{\DUtablewidth}{\linewidth}
\begin{longtable*}[c]{p{0.470\DUtablewidth}p{0.470\DUtablewidth}}

{\ttfamily \raggedright \noindent
\DUrole{k}{match}~\DUrole{n}{e}~\DUrole{k}{with}~\\
~\DUrole{o}{|}~\DUrole{mi}{0}~\DUrole{o}{->}~\DUrole{mi}{1}~\\
~\DUrole{o}{|}~\DUrole{n}{n}~\DUrole{k}{when}~\DUrole{n}{n}~\DUrole{o}{>}~\DUrole{mi}{10}~\DUrole{o}{->}~\DUrole{mi}{2}~\\
~\DUrole{o}{|}~\DUrole{o}{\_}~\DUrole{o}{->}~\DUrole{mi}{3}
}
 & 
{\ttfamily \raggedright \noindent
\DUrole{kr}{case}~\DUrole{n}{e}~\DUrole{kr}{of}~\\
~\DUrole{mi}{0}~\DUrole{ow}{->}~\DUrole{mi}{1}~\\
~\DUrole{n}{n}~\DUrole{o}{|}~\DUrole{n}{n}~\DUrole{o}{>}~\DUrole{mi}{10}~\DUrole{ow}{->}~\DUrole{mi}{2}~\\
~\DUrole{kr}{\_}~\DUrole{ow}{->}~\DUrole{mi}{3}
}
 \\
\end{longtable*}

Lists:

\setlength{\DUtablewidth}{\linewidth}
\begin{longtable*}[c]{p{0.470\DUtablewidth}p{0.470\DUtablewidth}}

{\ttfamily \raggedright \noindent
\DUrole{k}{let}~\DUrole{k}{rec}~\DUrole{n}{len}~\DUrole{o}{=}~\DUrole{k}{function}~\\
~~~\DUrole{o}{|}~\DUrole{bp}{{[}{]}}~\DUrole{o}{->}~\DUrole{mi}{0}~\\
~~~\DUrole{o}{|}~\DUrole{n}{x}\DUrole{o}{::}\DUrole{n}{xs}~\DUrole{o}{->}~\DUrole{mi}{1}~\DUrole{o}{+}~\DUrole{o}{(}\DUrole{n}{len}~\DUrole{n}{xs}\DUrole{o}{)}~\\
\DUrole{k}{let}~\DUrole{n}{v}~\DUrole{o}{=}~\DUrole{n}{len}~\DUrole{o}{({[}}\DUrole{mi}{1}\DUrole{o}{;}~\DUrole{mi}{2}\DUrole{o}{{]}}~\DUrole{o}{@}~\DUrole{o}{{[}}\DUrole{mi}{2}\DUrole{o}{;}~\DUrole{mi}{1}\DUrole{o}{{]}}~\DUrole{o}{@}~\DUrole{mi}{3}\DUrole{o}{::}\DUrole{mi}{4}\DUrole{o}{::}\DUrole{bp}{{[}{]}}\DUrole{o}{)}~\\
\DUrole{c}{(*~v~=~6~*)}
}
 & 
{\ttfamily \raggedright \noindent
\DUrole{nf}{len}~\DUrole{kt}{{[}{]}}~\DUrole{ow}{=}~\DUrole{mi}{0}~\\
\DUrole{nf}{len}~\DUrole{p}{(}\DUrole{n}{x}\DUrole{kt}{:}\DUrole{n}{xs}\DUrole{p}{)}~\DUrole{ow}{=}~\DUrole{mi}{1}~\DUrole{o}{+}~\DUrole{p}{(}\DUrole{n}{len}~\DUrole{n}{xs}\DUrole{p}{)}~\\
~\\
\DUrole{nf}{v}~\DUrole{ow}{=}~\DUrole{n}{len}~\DUrole{p}{({[}}\DUrole{mi}{1}\DUrole{p}{,}~\DUrole{mi}{2}\DUrole{p}{{]}}~\DUrole{o}{++}~\DUrole{p}{{[}}\DUrole{mi}{2}\DUrole{p}{,}~\DUrole{mi}{1}\DUrole{p}{{]}}~\DUrole{o}{++}~\DUrole{mi}{3}\DUrole{kt}{:}\DUrole{mi}{4}\DUrole{kt}{:{[}{]}}\DUrole{p}{)}~\\
\DUrole{cm}{\{-~v~=~6~-\}}
}
 \\
\end{longtable*}

Parametric types:

\setlength{\DUtablewidth}{\linewidth}
\begin{longtable*}[c]{p{0.470\DUtablewidth}p{0.470\DUtablewidth}}

{\ttfamily \raggedright \noindent
\DUrole{c}{(*~type~parameters~use~an~apostrophe~*)}~\\
\DUrole{c}{(*~tuple~types~defined~with~\textquotedbl{}*\textquotedbl{}~*)}~\\
\DUrole{k}{type}~\DUrole{k}{'}\DUrole{n}{a}~\DUrole{n}{pair}~\DUrole{o}{=}~\DUrole{nc}{P}~\DUrole{k}{of}~\DUrole{o}{(}\DUrole{k}{'}\DUrole{n}{a}~\DUrole{o}{*}~\DUrole{k}{'}\DUrole{n}{a}\DUrole{o}{)}~\\
~\\
\DUrole{c}{(*~concrete~parameter\\
~~~before~abstract~type~*)}~\\
\DUrole{k}{type}~\DUrole{n}{t}~\DUrole{o}{=}~\DUrole{kt}{int}~\DUrole{n}{pair}
}
 & 
{\ttfamily \raggedright \noindent
\DUrole{cm}{\{-~type~parameters~use~small~letters~-\}}~\\
\DUrole{cm}{\{-~tuple~types~defined~by~commas~-\}}~\\
\DUrole{kr}{data}~\DUrole{kt}{Pair}~\DUrole{n}{a}~\DUrole{ow}{=}~\DUrole{kt}{P}~\DUrole{p}{(}\DUrole{n}{a}\DUrole{p}{,}~\DUrole{n}{a}\DUrole{p}{)}~\\
~\\
\DUrole{cm}{\{-~concrete~parameter\\
~~~after~abstract~type~-\}}~\\
\DUrole{kr}{type}~\DUrole{kt}{T}~\DUrole{ow}{=}~\DUrole{kt}{Pair}~\DUrole{kt}{Int}
}
 \\
\end{longtable*}

Algebraic data types:

\setlength{\DUtablewidth}{\linewidth}
\begin{longtable*}[c]{p{0.470\DUtablewidth}p{0.470\DUtablewidth}}

{\ttfamily \raggedright \noindent
\DUrole{c}{(*~predefined~*)}~\\
\DUrole{k}{type}~\DUrole{k}{'}\DUrole{n}{a}~\DUrole{n}{option}~\DUrole{o}{=}~\\
~~~~~\DUrole{o}{|}~\DUrole{nc}{None}~\\
~~~~~\DUrole{o}{|}~\DUrole{nc}{Some}~\DUrole{k}{of}~\DUrole{k}{'}\DUrole{n}{a}~\\
~\\
\DUrole{c}{(*~custom~*)}~\\
\DUrole{k}{type}~\DUrole{k}{'}\DUrole{n}{a}~\DUrole{n}{tree}~\DUrole{o}{=}~\\
~~~~~\DUrole{o}{|}~\DUrole{nc}{Leaf}~\DUrole{k}{of}~\DUrole{k}{'}\DUrole{n}{a}~\\
~~~~~\DUrole{o}{|}~\DUrole{nc}{Node}~\DUrole{k}{of}~\DUrole{k}{'}\DUrole{n}{a}~\DUrole{n}{tree}~\DUrole{o}{*}~\DUrole{k}{'}\DUrole{n}{a}~\DUrole{n}{tree}
}
 & 
{\ttfamily \raggedright \noindent
\DUrole{cm}{\{-~predefined~-\}}~\\
\DUrole{kr}{data}~\DUrole{kt}{Maybe}~\DUrole{n}{a}~\DUrole{ow}{=}~\\
~~~~~\DUrole{o}{|}~\DUrole{kt}{Nothing}~\\
~~~~~\DUrole{o}{|}~\DUrole{kt}{Just}~\DUrole{n}{a}~\\
~\\
\DUrole{cm}{\{-~custom~-\}}~\\
\DUrole{kr}{data}~\DUrole{kt}{Tree}~\DUrole{n}{a}~\DUrole{ow}{=}~\\
~~~~~\DUrole{o}{|}~\DUrole{kt}{Leaf}~\DUrole{n}{a}~\\
~~~~~\DUrole{o}{|}~\DUrole{kt}{Node}~\DUrole{p}{(}\DUrole{kt}{Tree}~\DUrole{n}{a}\DUrole{p}{,}~\DUrole{kt}{Tree}~\DUrole{n}{a}\DUrole{p}{)}
}
 \\
\end{longtable*}

Generalized algebraic data types:

\setlength{\DUtablewidth}{\linewidth}
\begin{longtable*}[c]{p{0.470\DUtablewidth}p{0.470\DUtablewidth}}

{\ttfamily \raggedright \noindent
\DUrole{c}{(*~available~from~OCaml~4.x~*)}~\\
\DUrole{k}{type}~\DUrole{o}{\_}~\DUrole{n}{term}~\DUrole{o}{=}~\\
~~\DUrole{o}{|}~\DUrole{nc}{Int}~\DUrole{o}{:}~\DUrole{kt}{int}~\DUrole{o}{->}~\DUrole{kt}{int}~\DUrole{n}{term}~\\
~~\DUrole{o}{|}~\DUrole{nc}{Add}~\DUrole{o}{:}~\DUrole{o}{(}\DUrole{kt}{int}~\DUrole{o}{->}~\DUrole{kt}{int}~\DUrole{o}{->}~\DUrole{kt}{int}\DUrole{o}{)}~\DUrole{n}{term}~\\
~~\DUrole{o}{|}~\DUrole{nc}{App}~\DUrole{o}{:}~\DUrole{o}{(}\DUrole{k}{'}\DUrole{n}{b}~\DUrole{o}{->}~\DUrole{k}{'}\DUrole{n}{a}\DUrole{o}{)}~\DUrole{n}{term}~\DUrole{o}{*}~\DUrole{k}{'}\DUrole{n}{b}~\DUrole{n}{term}~\\
~~~~~~~~~~\DUrole{o}{->}~\DUrole{k}{'}\DUrole{n}{a}~\DUrole{n}{term}~\\
~\\
\DUrole{k}{let}~\DUrole{k}{rec}~\DUrole{n}{eval}~\DUrole{o}{:}~\DUrole{k}{type}~\DUrole{n}{a}\DUrole{o}{.}~\DUrole{n}{a}~\DUrole{n}{term}~\DUrole{o}{->}~\DUrole{n}{a}~\DUrole{o}{=}~\\
\DUrole{k}{function}~\\
~~\DUrole{o}{|}~\DUrole{nc}{Int}~\DUrole{n}{n}~~~~\DUrole{o}{->}~\DUrole{n}{n}~\\
~~\DUrole{o}{|}~\DUrole{nc}{Add}~~~~~~\DUrole{o}{->}~\DUrole{o}{(}\DUrole{k}{fun}~\DUrole{n}{x}~\DUrole{n}{y}~\DUrole{o}{->}~\DUrole{n}{x}\DUrole{o}{+}\DUrole{n}{y}\DUrole{o}{)}~\\
~~\DUrole{o}{|}~\DUrole{nc}{App}\DUrole{o}{(}\DUrole{n}{f}\DUrole{o}{,}\DUrole{n}{x}\DUrole{o}{)}~\DUrole{o}{->}~\DUrole{o}{(}\DUrole{n}{eval}~\DUrole{n}{f}\DUrole{o}{)}~\DUrole{o}{(}\DUrole{n}{eval}~\DUrole{n}{x}\DUrole{o}{)}~\\
~\\
\DUrole{c}{(*~two~:~int~*)}~\\
\DUrole{k}{let}~\DUrole{n}{two}~\DUrole{o}{=}~\\
~~~~\DUrole{n}{eval}~\DUrole{o}{(}\DUrole{nc}{App}~\DUrole{o}{(}\DUrole{nc}{App}~\DUrole{o}{(}\DUrole{nc}{Add}\DUrole{o}{,}~\DUrole{nc}{Int}~\DUrole{mi}{1}\DUrole{o}{),}~\DUrole{nc}{Int}~\DUrole{mi}{1}\DUrole{o}{))}
}
 & 
{\ttfamily \raggedright \noindent
\DUrole{cm}{\{-~compile~with~-XGADTs~-\}}~\\
\DUrole{kr}{data}~\DUrole{kt}{Term}~\DUrole{n}{a}~\DUrole{kr}{where}~\\
~~\DUrole{kt}{Int}~\DUrole{ow}{::}~\DUrole{kt}{Int}~\DUrole{ow}{->}~\DUrole{kt}{Term}~\DUrole{kt}{Int}~\\
~~\DUrole{kt}{Add}~\DUrole{ow}{::}~\DUrole{kt}{Term}~\DUrole{p}{(}\DUrole{kt}{Int}~\DUrole{ow}{->}~\DUrole{kt}{Int}~\DUrole{ow}{->}~\DUrole{kt}{Int}\DUrole{p}{)}~\\
~~\DUrole{kt}{App}~\DUrole{ow}{::}~\DUrole{p}{(}\DUrole{kt}{Term}~\DUrole{p}{(}\DUrole{n}{b}~\DUrole{ow}{->}~\DUrole{n}{a}\DUrole{p}{),}~\DUrole{kt}{Term}~\DUrole{n}{b}\DUrole{p}{)}~\\
~~~~~~~~~\DUrole{ow}{->}~\DUrole{kt}{Term}~\DUrole{n}{a}~\\
~\\
\DUrole{nf}{eval}~\DUrole{ow}{::}~\DUrole{kt}{Term}~\DUrole{n}{a}~\DUrole{ow}{->}~\DUrole{n}{a}~\\
~\\
\DUrole{nf}{eval}~\DUrole{p}{(}\DUrole{kt}{Int}~\DUrole{n}{n}\DUrole{p}{)}~~~~\DUrole{ow}{=}~\DUrole{n}{n}~\\
\DUrole{nf}{eval}~\DUrole{p}{(}\DUrole{kt}{Add}\DUrole{p}{)}~~~~~~\DUrole{ow}{=}~\DUrole{p}{(}\DUrole{nf}{\textbackslash{}}\DUrole{n}{x}~\DUrole{n}{y}~\DUrole{ow}{->}~\DUrole{n}{x}~\DUrole{o}{+}~\DUrole{n}{y}\DUrole{p}{)}~\\
\DUrole{nf}{eval}~\DUrole{p}{(}\DUrole{kt}{App}\DUrole{p}{(}\DUrole{n}{f}\DUrole{p}{,}\DUrole{n}{x}\DUrole{p}{))}~\DUrole{ow}{=}~\DUrole{p}{(}\DUrole{n}{eval}~\DUrole{n}{f}\DUrole{p}{)}~\DUrole{p}{(}\DUrole{n}{eval}~\DUrole{n}{x}\DUrole{p}{)}~\\
~\\
\DUrole{nf}{two}~\DUrole{ow}{::}~\DUrole{kt}{Int}~\\
\DUrole{nf}{two}~\DUrole{ow}{=}~\\
~~~~\DUrole{n}{eval}~\DUrole{p}{(}\DUrole{kt}{App}~\DUrole{p}{(}\DUrole{kt}{App}~\DUrole{p}{(}\DUrole{kt}{Add}\DUrole{p}{,}~\DUrole{kt}{Int}~\DUrole{mi}{1}\DUrole{p}{),}~\DUrole{kt}{Int}~\DUrole{mi}{1}\DUrole{p}{))}
}
 \\
\end{longtable*}

Text I/O, string representation and interpretation (yay to overloading):
\begin{quote}{\ttfamily \raggedright \noindent
string\_of\_int~string\_of\_float~~~~~~~~~~(*~OCaml~*)\\
show~~~~~~~~~~show~~~~~~~~~~~~~~~~~~~~~\{-~Haskell~-\}\\
~\\
int\_of\_string~float\_of\_string~~~~~~~~~~(*~OCaml~*)\\
read~~~~~~~~~~read~~~~~~~~~~~~~~~~~~~~~\{-~Haskell~-\}\\
~\\
print\_char~print\_string~print\_endline~~(*~OCaml~*)\\
putChar~~~~putStr~~~~~~~putStrLn~~~~~~~\{-~Haskell~-\}\\
~\\
input\_line~input\_char~~~~~~~~~~~~~~~~~~(*~OCaml~*)\\
getLine~~~~getChar~~~~~~~~~~~~~~~~~~~~~\{-~Haskell~-\}
}
\end{quote}

Functional goodies:

\setlength{\DUtablewidth}{\linewidth}
\begin{longtable*}[c]{p{0.470\DUtablewidth}p{0.470\DUtablewidth}}

{\ttfamily \raggedright \noindent
\DUrole{c}{(*~\textquotedbl{}@@\textquotedbl{}~is~low-priority~application~*)}~\\
\DUrole{n}{print\_endline}~\DUrole{o}{@@}~\DUrole{n}{string\_of\_int}~\DUrole{mi}{3}~\\
~\\
\DUrole{c}{(*~can~be~defined~in~OCaml~*)}~\\
\DUrole{k}{val}~\DUrole{n}{flip}~\DUrole{o}{:}~\DUrole{o}{(}\DUrole{k}{'}\DUrole{n}{a}~\DUrole{o}{->}~\DUrole{k}{'}\DUrole{n}{b}~\DUrole{o}{->}~\DUrole{k}{'}\DUrole{n}{c}\DUrole{o}{)}~\\
~~~~~~~~~~~\DUrole{o}{->}~\DUrole{k}{'}\DUrole{n}{b}~\DUrole{o}{->}~\DUrole{k}{'}\DUrole{n}{a}~\DUrole{o}{->}~\DUrole{k}{'}\DUrole{n}{c}~\\
\DUrole{k}{let}~\DUrole{n}{flip}~\DUrole{n}{f}~\DUrole{o}{=}~\DUrole{k}{fun}~\DUrole{n}{x}~\DUrole{n}{y}~\DUrole{o}{->}~\DUrole{n}{f}~\DUrole{n}{y}~\DUrole{n}{x}~\\
~\\
\DUrole{k}{val}~\DUrole{o}{(@.)}~\DUrole{o}{:}~\DUrole{o}{(}\DUrole{k}{'}\DUrole{n}{b}~\DUrole{o}{->}~\DUrole{k}{'}\DUrole{n}{c}\DUrole{o}{)}~\DUrole{o}{->}~\DUrole{o}{(}\DUrole{k}{'}\DUrole{n}{a}~\DUrole{o}{->}~\DUrole{k}{'}\DUrole{n}{b}\DUrole{o}{)}~\\
~~~~~~~~~~~\DUrole{o}{->}~\DUrole{k}{'}\DUrole{n}{a}~\DUrole{o}{->}~\DUrole{k}{'}\DUrole{n}{c}~\\
\DUrole{k}{let}~\DUrole{o}{(@.)}~\DUrole{n}{f}~\DUrole{n}{g}~\DUrole{o}{=}~\DUrole{k}{fun}~\DUrole{n}{x}~\DUrole{o}{->}~\DUrole{n}{f}~\DUrole{o}{(}\DUrole{n}{g}~\DUrole{n}{x}\DUrole{o}{)}
}
 & 
{\ttfamily \raggedright \noindent
\DUrole{cm}{\{-~\textquotedbl{}\$\textquotedbl{}~is~low-priority~application~-\}}~\\
\DUrole{nf}{putStrLn}~\DUrole{o}{\$}~\DUrole{n}{show}~\DUrole{mi}{3}~\\
~\\
\DUrole{cm}{\{-~built-in~in~Haskell~-\}}~\\
\DUrole{nf}{flip}~\DUrole{ow}{::}~\DUrole{p}{(}\DUrole{n}{a}~\DUrole{ow}{->}~\DUrole{n}{b}~\DUrole{ow}{->}~\DUrole{n}{c}\DUrole{p}{)}~\\
~~~~~~~~\DUrole{ow}{->}~\DUrole{n}{b}~\DUrole{ow}{->}~\DUrole{n}{a}~\DUrole{ow}{->}~\DUrole{n}{c}~\\
\DUrole{nf}{flip}~\DUrole{n}{f}~\DUrole{ow}{=}~\DUrole{nf}{\textbackslash{}}\DUrole{n}{x}~\DUrole{n}{y}~\DUrole{ow}{->}~\DUrole{n}{f}~\DUrole{n}{y}~\DUrole{n}{x}~\\
~\\
\DUrole{p}{(}\DUrole{o}{.}\DUrole{p}{)}~\DUrole{ow}{::}~\DUrole{p}{(}\DUrole{n}{b}~\DUrole{ow}{->}~\DUrole{n}{c}\DUrole{p}{)}~\DUrole{ow}{->}~\DUrole{p}{(}\DUrole{n}{a}~\DUrole{ow}{->}~\DUrole{n}{b}\DUrole{p}{)}~\\
~~~~~~~\DUrole{ow}{->}~\DUrole{n}{a}~\DUrole{ow}{->}~\DUrole{n}{c}~\\
\DUrole{p}{(}\DUrole{o}{.}\DUrole{p}{)}~\DUrole{n}{f}~\DUrole{n}{g}~\DUrole{ow}{=}~\DUrole{nf}{\textbackslash{}}\DUrole{n}{x}~\DUrole{ow}{->}~\DUrole{n}{f}~\DUrole{p}{(}\DUrole{n}{g}~\DUrole{n}{x}\DUrole{p}{)}
}
 \\
\end{longtable*}

\DUtransition

\section{Recursive definitions%
  \label{recursive-definitions}%
}

In OCaml, recursive definitions are expressed with the keyword
\texttt{rec}. In Haskell, \emph{all definitions are recursive}, including those
of variables. This means that one cannot use the same identifier to
bind to successive definitions in the same scope:

\setlength{\DUtablewidth}{\linewidth}
\begin{longtable*}[c]{p{0.470\DUtablewidth}p{0.470\DUtablewidth}}

{\ttfamily \raggedright \noindent
\DUrole{k}{let}~\DUrole{n}{f}~\DUrole{n}{n}~\DUrole{o}{=}~\\
~~~~\DUrole{c}{(*~this~defines~a~fresh~\textquotedbl{}n\textquotedbl{}~*)}~\\
~~~~\DUrole{k}{let}~\DUrole{n}{n}~\DUrole{o}{=}~\DUrole{n}{n}~\DUrole{o}{-}~\DUrole{mi}{1}~\DUrole{k}{in}~\\
~~~~\DUrole{n}{n}
}
 & 
{\ttfamily \raggedright \noindent
\DUrole{nf}{f}~\DUrole{n}{n}~\DUrole{ow}{=}~\\
~~~~\DUrole{cm}{\{-~must~use~different~identifiers~-\}}~\\
~~~~\DUrole{kr}{let}~\DUrole{n}{n'}~\DUrole{ow}{=}~\DUrole{n}{n}~\DUrole{o}{-}~\DUrole{mi}{1}~\DUrole{kr}{in}~\\
~~~~\DUrole{n}{n'}
}
 \\
\end{longtable*}

Otherwise, the Haskell compiler will generate a circular data definition
for \texttt{n}, which is usually not the intended result!

\section{Program structure%
  \label{program-structure}%
}

In OCaml, a \emph{program} is defined by the concatenation of all the
statements in the source code; the control entry point at run-time is
the first statement in the top-level module. There is no function with
a special role as entry point. Execution proceeds by evaluating all
statements in order.
\begin{quote}{\ttfamily \raggedright \noindent
\DUrole{k}{let}~\DUrole{n}{f}~\DUrole{o}{=}~\\
~~~~\DUrole{n}{print\_endline}~\DUrole{s2}{\textquotedbl{}start\textquotedbl{}}\DUrole{o}{;}~\\
~~~~\DUrole{k}{fun}~\DUrole{bp}{()}~\DUrole{o}{->}~\DUrole{s2}{\textquotedbl{}world\textquotedbl{}}~\\
~\\
\DUrole{k}{let}~\DUrole{n}{p1}~\DUrole{bp}{()}~\DUrole{o}{=}~\\
~~\DUrole{n}{print\_endline}~\DUrole{s2}{\textquotedbl{}hello\textquotedbl{}}~\\
\DUrole{k}{let}~\DUrole{n}{p2}~\DUrole{bp}{()}~\DUrole{o}{=}~\\
~~\DUrole{n}{print\_endline}~\DUrole{o}{(}\DUrole{n}{f}~\DUrole{bp}{()}\DUrole{o}{)}~\\
~\\
\DUrole{n}{p1}~\DUrole{bp}{()}\DUrole{o}{;}~\DUrole{n}{p2}~\DUrole{bp}{()}~\\
~\\
\DUrole{c}{(*~prints~\textquotedbl{}start\textquotedbl{},~\textquotedbl{}hello\textquotedbl{},~\textquotedbl{}world\textquotedbl{}~*)}
}
\end{quote}

In Haskell, a “program” must be defined by a \emph{constant} (not function)
with the special name “\textbf{main}”. The run-time system starts execution
by first constructing the data representations of all top-level
definitions; including the definition of the constant “main”, which
stores a \emph{data structure} akin to a “list of statements”, or
“recipe”. At that point, there cannot be any effects performed
yet. Then the run-time system continues execution by reducing the
recipe referenced by main, step by step from the head, performing the
effects described by the statements therein.

To build such a “list of statements” or “recipe”, Haskell provides a shorthand
syntax using the \textbf{do} keyword and newline/semicolon separators:
\begin{quote}{\ttfamily \raggedright \noindent
\DUrole{cm}{\{-~no~side-effects~in~top-level~definitions~-\}}~\\
\DUrole{nf}{f}~\DUrole{nb}{()}~\DUrole{ow}{=}~\DUrole{s}{\textquotedbl{}world\textquotedbl{}}~\\
~\\
\DUrole{nf}{p1}~\DUrole{ow}{=}~\DUrole{kr}{do}~\DUrole{cm}{\{-~p1~is~a~constant~recipe~-\}}~\\
~~\DUrole{n}{putStrLn}~\DUrole{s}{\textquotedbl{}hello\textquotedbl{}}~\\
\DUrole{nf}{p2}~\DUrole{ow}{=}~\DUrole{kr}{do}~\DUrole{cm}{\{-~p2~is~another~constant~recipe~-\}}~\\
~~\DUrole{n}{putStrLn}~\DUrole{p}{(}\DUrole{n}{f}~\DUrole{nb}{()}\DUrole{p}{)}~\\
~\\
\DUrole{cm}{\{-~entry~program~must~be~called~\textquotedbl{}main\textquotedbl{}~-\}}~\\
\DUrole{cm}{\{-~it~is~also~a~constant~recipe~-\}}~\\
\DUrole{nf}{main}~\DUrole{ow}{=}~\DUrole{kr}{do}~\DUrole{n}{p1}\DUrole{p}{;}~\DUrole{n}{p2}
}
\end{quote}

In OCaml, all statements
in source files are executed when the corresponding modules are
imported. In Haskell, source code files provide only definitions;
there are no side-effects performed during module imports.

\setlength{\DUtablewidth}{\linewidth}
\begin{longtable*}[c]{p{0.470\DUtablewidth}p{0.470\DUtablewidth}}

{\ttfamily \raggedright \noindent
\DUrole{c}{(*~in~t.ml~*)}~\\
~\\
\DUrole{k}{let}~\DUrole{o}{\_}~\DUrole{o}{=}~\\
~~~~\DUrole{n}{print\_endline}~\DUrole{s2}{\textquotedbl{}hello\textquotedbl{}}~\\
~\\
\DUrole{c}{(*~in~main.ml~*)}~\\
\DUrole{k}{open}~\DUrole{nc}{T}~\\
\DUrole{k}{let}~\DUrole{n}{main}~\DUrole{o}{=}~\\
~~~~\DUrole{n}{print\_endline}~\DUrole{s2}{\textquotedbl{}world\textquotedbl{}}~\\
~\\
\DUrole{c}{(*~prints~\textquotedbl{}hello\textquotedbl{},~then~\textquotedbl{}world\textquotedbl{}~*)}
}
 & 
{\ttfamily \raggedright \noindent
\DUrole{cm}{\{-~in~T.hs~-\}}~\\
\DUrole{kr}{module}~\DUrole{nn}{T}~\DUrole{kr}{where}~\\
\DUrole{kr}{\_}~\DUrole{ow}{=}~\DUrole{kr}{do}~\\
~~~~\DUrole{n}{putStrLn}~\DUrole{s}{\textquotedbl{}hello\textquotedbl{}}~\\
~\\
\DUrole{cm}{\{-~in~main.hs~-\}}~\\
\DUrole{kr}{import}~\DUrole{nn}{T}~\\
\DUrole{nf}{main}~\DUrole{ow}{=}~\DUrole{kr}{do}~\\
~~~~\DUrole{n}{putStrLn}~\DUrole{s}{\textquotedbl{}world\textquotedbl{}}~\\
~\\
\DUrole{cm}{\{-~prints~only~\textquotedbl{}world\textquotedbl{},~not~\textquotedbl{}hello\textquotedbl{}~-\}}
}
 \\
\end{longtable*}

This is because only the constant recipe constructed by the “main”
definition will be evaluated for side-effects at run-time, after all
top-level definitions are constructed.

In OCaml, a file named \textquotedbl{}\texttt{blah.ml}\textquotedbl{} defines a module
named \textquotedbl{}Blah\textquotedbl{}. The module name is derived from the file name.
In Haskell, a \textbf{module} specification in the file
specifies the module name:
\begin{quote}{\ttfamily \raggedright \noindent
\DUrole{kr}{module}~\DUrole{nn}{Blah}~\DUrole{kr}{where}~\DUrole{o}{...}
}
\end{quote}

\emph{By convention}, programmers also name the file containing \textquotedbl{}\texttt{module
Blah where ...}\textquotedbl{} with filename \texttt{Blah.hs}. However, it is possible
to split a Haskell module definition in multiple source files with
different names, or define multiple Haskell modules in the same source
file\DUfootnotemark{id4}{id5}{1}.
\DUfootnotetext{id5}{id4}{1}{%
This opportunity is specified in the \href{http://www.haskell.org/onlinereport/modules.html}{Haskell Language
Report}, the official specification of the Haskell
language. However, in practice, its most popular implementation GHC
only properly handles modules where the filename matches the
\texttt{module} specification therein.
}

Module use:

\setlength{\DUtablewidth}{\linewidth}
\begin{longtable*}[c]{p{0.470\DUtablewidth}p{0.470\DUtablewidth}}

{\ttfamily \raggedright \noindent
\DUrole{k}{open}~\DUrole{nc}{Char}~\\
~\\
\DUrole{c}{(*~\textquotedbl{}lowercase\textquotedbl{}~imported~by\\
~~~\textquotedbl{}open\textquotedbl{}~from~Char~*)}~\\
\DUrole{k}{let}~\DUrole{n}{lc}~\DUrole{o}{=}~\DUrole{n}{lowercase}~\\
~\\
\DUrole{c}{(*~no~explicit~import~needed\\
~~~to~use~qualified~module~names~*)}~\\
\DUrole{k}{let}~\DUrole{n}{lsz}~~\DUrole{o}{=}~\DUrole{nn}{List}\DUrole{p}{.}\DUrole{n}{length}
}
 & 
{\ttfamily \raggedright \noindent
\DUrole{kr}{import}~\DUrole{nn}{Char}~\\
~\\
\DUrole{cm}{\{-~\textquotedbl{}toLower\textquotedbl{}~imported~by\\
~~~\textquotedbl{}import\textquotedbl{}~from~Char~-\}}~\\
\DUrole{nf}{lc}~\DUrole{ow}{=}~\DUrole{n}{toLower}~\\
~\\
\DUrole{cm}{\{-~need~\textquotedbl{}import~qualified\textquotedbl{}~to~use~-\}}~\\
\DUrole{kr}{import}~\DUrole{k}{qualified}~\DUrole{nn}{Data.List}~\\
\DUrole{nf}{lsz}~\DUrole{ow}{=}~\DUrole{kt}{Data}\DUrole{o}{.}\DUrole{kt}{List}\DUrole{o}{.}\DUrole{n}{length}
}
 \\
\end{longtable*}

It is possible to import a module in Haskell (OCaml's “open”) without
importing all the names it defines. To include only specific
identifiers, place them between parentheses after the module name; to
exclude specific identifiers, use “\texttt{hiding}”:
\begin{quote}{\ttfamily \raggedright \noindent
\DUrole{kr}{import}~\DUrole{nn}{Char}~\DUrole{p}{(}\DUrole{nf}{toUpper}\DUrole{p}{,}\DUrole{nf}{toLower}\DUrole{p}{)}~\\
\DUrole{kr}{import}~\DUrole{nn}{Char}~\DUrole{k}{hiding}~\DUrole{p}{(}\DUrole{nf}{isAscii}\DUrole{p}{)}
}
\end{quote}

The set of predefined functions and operators in OCaml is provided by
the module \texttt{Pervasives}. In Haskell, they are provided by module
\texttt{Prelude}. It is possible to disable predefined functions using
\texttt{import Prelude hiding (...)}.

Contrary to OCaml, it is not possible in Haskell to import/open a module locally to a
function. Also, there are no direct Haskell equivalents for OCaml's parametric modules
(functors). See \hyperref[haskell-limitations]{Haskell limitations} below for suggestions.

\section{Program and effect composition%
  \label{program-and-effect-composition}%
}

Composition of pure functions in Haskell is conceptually the same as
in OCaml. OCaml predefines \textquotedbl{}\texttt{|>}\textquotedbl{} (reverse pipeline), Haskell predefines \textquotedbl{}\texttt{.}\textquotedbl{} (composition).

\setlength{\DUtablewidth}{\linewidth}
\begin{longtable*}[c]{p{0.470\DUtablewidth}p{0.470\DUtablewidth}}

{\ttfamily \raggedright \noindent
\DUrole{k}{let}~\DUrole{n}{p}~\DUrole{n}{x}~\DUrole{o}{=}~\DUrole{n}{f}~\DUrole{o}{(}\DUrole{n}{g}~\DUrole{o}{(}\DUrole{n}{h}~\DUrole{n}{x}\DUrole{o}{))}~\\
\DUrole{k}{let}~\DUrole{n}{p}~\DUrole{n}{x}~\DUrole{o}{=}~\DUrole{n}{f}~\DUrole{o}{@@}~\DUrole{n}{g}~\DUrole{o}{@@}~\DUrole{n}{h}~\DUrole{o}{@@}~\DUrole{n}{x}~\\
~\\
\DUrole{c}{(*~\textquotedbl{}|>\textquotedbl{}~predefined~in~OCaml~*)}~\\
~\\
\DUrole{k}{let}~\DUrole{n}{p}~\DUrole{n}{x}~\DUrole{o}{=}~\DUrole{n}{x}~\DUrole{o}{|>}~\DUrole{n}{h}~\DUrole{o}{|>}~\DUrole{n}{g}~\DUrole{o}{|>}~\DUrole{n}{f}~\\
~\\
\DUrole{c}{(*~\textquotedbl{}.\textquotedbl{}~not~predefined~in~OCaml~*)}~\\
\DUrole{k}{let}~\DUrole{o}{(@.)}~\DUrole{n}{f}~\DUrole{n}{g}~\DUrole{o}{=}~\DUrole{k}{fun}~\DUrole{n}{x}~\DUrole{o}{->}~\DUrole{n}{f}~\DUrole{o}{(}\DUrole{n}{g}~\DUrole{n}{x}\DUrole{o}{)}~\\
\DUrole{k}{let}~\DUrole{n}{p}~\DUrole{n}{x}~\DUrole{o}{=}~\DUrole{o}{(}\DUrole{n}{f}~\DUrole{o}{@.}~\DUrole{n}{g}~\DUrole{o}{@.}~\DUrole{n}{h}\DUrole{o}{)}~\DUrole{n}{x}
}
 & 
{\ttfamily \raggedright \noindent
\DUrole{nf}{p}~\DUrole{n}{x}~\DUrole{ow}{=}~\DUrole{n}{f}~\DUrole{p}{(}\DUrole{n}{g}~\DUrole{p}{(}\DUrole{n}{h}~\DUrole{n}{x}\DUrole{p}{))}~\\
\DUrole{nf}{p}~\DUrole{n}{x}~\DUrole{ow}{=}~\DUrole{n}{f}~\DUrole{o}{\$}~\DUrole{n}{g}~\DUrole{o}{\$}~\DUrole{n}{h}~\DUrole{o}{\$}~\DUrole{n}{x}~\\
~\\
\DUrole{cm}{\{-~\textquotedbl{}|>\textquotedbl{}~not~predefined~in~Haskell~-\}}~\\
\DUrole{p}{(}\DUrole{o}{|>}\DUrole{p}{)}~\DUrole{n}{x}~\DUrole{n}{f}~\DUrole{ow}{=}~\DUrole{n}{f}~\DUrole{n}{x}~\\
\DUrole{nf}{p}~\DUrole{n}{x}~\DUrole{ow}{=}~\DUrole{n}{x}~\DUrole{o}{|>}~\DUrole{n}{h}~\DUrole{o}{|>}~\DUrole{n}{g}~\DUrole{o}{|>}~\DUrole{n}{f}~\\
~\\
\DUrole{cm}{\{-~\textquotedbl{}.\textquotedbl{}~predefined~in~Haskell~-\}}~\\
~\\
\DUrole{nf}{p}~\DUrole{n}{x}~\DUrole{ow}{=}~\DUrole{p}{(}\DUrole{n}{f}~\DUrole{o}{.}~\DUrole{n}{g}~\DUrole{o}{.}~\DUrole{n}{h}\DUrole{p}{)}~\DUrole{n}{x}
}
 \\
\end{longtable*}

To compose effectful statements with no result, both OCaml and Haskell
use semicolons (for Haskell, inside a do-block). Additionally in Haskell, a newline
character is also a valid statement separator in a do-block.

\setlength{\DUtablewidth}{\linewidth}
\begin{longtable*}[c]{p{0.470\DUtablewidth}p{0.470\DUtablewidth}}

{\ttfamily \raggedright \noindent
\DUrole{k}{val}~\DUrole{n}{f}~\DUrole{o}{:}~\DUrole{kt}{int}~\DUrole{o}{->}~\DUrole{kt}{unit}~\\
\DUrole{k}{val}~\DUrole{n}{g}~\DUrole{o}{:}~\DUrole{kt}{int}~\DUrole{o}{->}~\DUrole{kt}{unit}~\\
~\\
\DUrole{k}{let}~\DUrole{o}{\_}~\DUrole{o}{=}~\\
~~~~\DUrole{c}{(*~sequence~of~unit~calls~*)}~\\
~~~~\DUrole{n}{f}~\DUrole{mi}{3}~\DUrole{o}{;}~\DUrole{n}{f}~\DUrole{mi}{4}~\DUrole{o}{;}~\\
~~~~\DUrole{n}{g}~\DUrole{mi}{2}
}
 & 
{\ttfamily \raggedright \noindent
\DUrole{nf}{f}~\DUrole{ow}{::}~\DUrole{kt}{Int}~\DUrole{ow}{->}~\DUrole{kt}{IO}~\DUrole{nb}{()}~\\
\DUrole{nf}{g}~\DUrole{ow}{::}~\DUrole{kt}{Int}~\DUrole{ow}{->}~\DUrole{kt}{IO}~\DUrole{nb}{()}~\\
~\\
\DUrole{nf}{main}~\DUrole{ow}{=}~\DUrole{kr}{do}~\\
~~~~\DUrole{cm}{\{-~sequence~of~unit~statements~-\}}~\\
~~~~\DUrole{n}{f}~\DUrole{mi}{3}~\DUrole{p}{;}~\DUrole{n}{f}~\DUrole{mi}{4}~\DUrole{cm}{\{-~\textbackslash{}n~separates~too~-\}}~\\
~~~~\DUrole{n}{g}~\DUrole{mi}{2}
}
 \\
\end{longtable*}

In OCaml, effectful functions that return a value of type \texttt{t} are
declared with \texttt{t} as return type.

In Haskell, there are no effectful functions; only functions that
return “lists of statements”, or recipes, as described earlier. This creates a
conceptual level of indirection: there are no “functions that perform
an effect and return a value of type \texttt{t}”, but rather “functions
that return recipes, so that the subsequent evaluation of those recipes performs
effects and also produces a value of type \texttt{t}”.

A function that returns a recipe is typed with
return type “\texttt{IO t}”, which means “a recipe of statements that computes
a value of type \texttt{t} upon its later evaluation”.

To express the production of the computed value inside the do-block,
you can use the handy function \textbf{return}:

\setlength{\DUtablewidth}{\linewidth}
\begin{longtable*}[c]{p{0.470\DUtablewidth}p{0.470\DUtablewidth}}

{\ttfamily \raggedright \noindent
\DUrole{c}{(*~inc~:~int~->~int~~(with~effects)~*)}~\\
\DUrole{k}{let}~\DUrole{n}{inc}~\DUrole{n}{n}~\DUrole{o}{=}~\\
~~~~\DUrole{n}{print\_endline}~\DUrole{o}{@@}~\DUrole{n}{string\_of\_int}~\DUrole{n}{n}\DUrole{o}{;}~\\
~~~~\DUrole{n}{n}~\DUrole{o}{+}~\DUrole{mi}{1}~\\
~\\
\DUrole{c}{(*~fact~:~int~->~int~~(with~effects)~*)}~\\
\DUrole{k}{let}~\DUrole{n}{fact}~\DUrole{o}{=}~\\
~~~~\DUrole{k}{let}~\DUrole{k}{rec}~\DUrole{n}{factr}~\DUrole{n}{r}~\DUrole{n}{n}~\DUrole{o}{=}~\\
~~~~~~~\DUrole{n}{print\_endline}~\DUrole{o}{@@}~\DUrole{n}{string\_of\_int}~\DUrole{n}{n}\DUrole{o}{;}~\\
~~~~~~~\DUrole{k}{if}~\DUrole{n}{n}~\DUrole{o}{=}~\DUrole{mi}{1}~\DUrole{k}{then}~\DUrole{n}{r}~\\
~~~~~~~\DUrole{k}{else}~\DUrole{n}{factr}~\DUrole{o}{(}\DUrole{n}{n}~\DUrole{o}{*}~\DUrole{n}{r}\DUrole{o}{)}~\DUrole{o}{(}\DUrole{n}{n}~\DUrole{o}{-}~\DUrole{mi}{1}\DUrole{o}{)}~\\
~~~~\DUrole{k}{in}~\DUrole{n}{factr}~\DUrole{mi}{1}
}
 & 
{\ttfamily \raggedright \noindent
\DUrole{nf}{inc}~\DUrole{ow}{::}~\DUrole{kt}{Int}~\DUrole{ow}{->}~\DUrole{kt}{IO}~\DUrole{kt}{Int}~\\
\DUrole{nf}{inc}~\DUrole{n}{n}~\DUrole{ow}{=}~\DUrole{kr}{do}~\\
~~~~~~~\DUrole{n}{putStrLn}~\DUrole{o}{\$}~\DUrole{n}{show}~\DUrole{n}{n}~\\
~~~~~~~\DUrole{n}{return}~\DUrole{p}{(}\DUrole{n}{n}~\DUrole{o}{+}~\DUrole{mi}{1}\DUrole{p}{)}~\\
~\\
\DUrole{nf}{fact}~\DUrole{ow}{::}~\DUrole{kt}{Int}~\DUrole{ow}{->}~\DUrole{kt}{IO}~\DUrole{kt}{Int}~\\
\DUrole{nf}{fact}~\DUrole{ow}{=}~\\
~~~~\DUrole{kr}{let}~\DUrole{n}{factr}~\DUrole{n}{r}~\DUrole{n}{n}~\DUrole{ow}{=}~\DUrole{kr}{do}~\\
~~~~~~~\DUrole{n}{putStrLn}~\DUrole{o}{\$}~\DUrole{n}{show}~\DUrole{n}{n}~\\
~~~~~~~\DUrole{kr}{if}~\DUrole{n}{n}~\DUrole{o}{==}~\DUrole{mi}{1}~\DUrole{kr}{then}~\DUrole{n}{return}~\DUrole{n}{r}~\\
~~~~~~~\DUrole{kr}{else}~\DUrole{n}{factr}~\DUrole{p}{(}\DUrole{n}{n}~\DUrole{o}{*}~\DUrole{n}{r}\DUrole{p}{)}~\DUrole{p}{(}\DUrole{n}{n}~\DUrole{o}{-}~\DUrole{mi}{1}\DUrole{p}{)}~\\
~~~~\DUrole{kr}{in}~\DUrole{n}{factr}~\DUrole{mi}{1}
}
 \\
\end{longtable*}

In OCaml, you can bind a variable to the return value
of an effectful function with “\texttt{let}”, as usual. In Haskell,
another syntax is defined for this purpose using “\texttt{<-}” in a do-block:

\setlength{\DUtablewidth}{\linewidth}
\begin{longtable*}[c]{p{0.470\DUtablewidth}p{0.470\DUtablewidth}}

{\ttfamily \raggedright \noindent
\DUrole{c}{(*~fact~:~int~->~int~~(with~effects)~*)}~\\
~\\
\DUrole{k}{let}~\DUrole{o}{\_}~\DUrole{o}{=}~\\
~~~~\DUrole{k}{let}~\DUrole{n}{v}~\DUrole{o}{=}~\DUrole{n}{fact}~\DUrole{mi}{3}\DUrole{o}{;}~\\
~~~~\DUrole{n}{print\_endline}~\DUrole{o}{@@}~\DUrole{n}{string\_of\_int}~\DUrole{n}{v}~\\
~\\
\DUrole{c}{(*~prints~6~*)}
}
 & 
{\ttfamily \raggedright \noindent
\DUrole{nf}{fact}~\DUrole{ow}{::}~\DUrole{kt}{Int}~\DUrole{ow}{->}~\DUrole{kt}{IO}~\DUrole{kt}{Int}~\\
~\\
\DUrole{nf}{main}~\DUrole{ow}{=}~\DUrole{kr}{do}~\\
~~~~\DUrole{n}{v}~\DUrole{ow}{<-}~\DUrole{n}{fact}~\DUrole{mi}{3}~\\
~~~~\DUrole{n}{putStrLn}~\DUrole{o}{\$}~\DUrole{n}{show}~\DUrole{n}{v}~\\
~\\
\DUrole{cm}{\{-~prints~6~-\}}
}
 \\
\end{longtable*}

Haskell's insistence on separating pure functions from “lists of
effectful statements that produce values at run-time” creates a
plumbing problem that does not exist in OCaml. Say, for example, you
have two effectful functions like \texttt{inc} and \texttt{fact} above, that
both take a \emph{value} as argument and print something before producing
their result. How to compose them together?

Direct composition works in OCaml (because both return \texttt{int}, which
matches their input argument type), but not in Haskell. Instead in
Haskell we must either explicitly bind the return value of the first
effectful definition to a name with “\texttt{<-}”, or use the operators
“\texttt{=<{}<}” and “\texttt{>{}>=}” (piping of effects):

\setlength{\DUtablewidth}{\linewidth}
\begin{longtable*}[c]{p{0.470\DUtablewidth}p{0.470\DUtablewidth}}

{\ttfamily \raggedright \noindent
\DUrole{c}{(*~inc~:~int~->~int~~~(with~effects)~*)}~\\
\DUrole{c}{(*~fact~:~int~->~int~~(with~effects)~*)}~\\
~\\
\DUrole{k}{let}~\DUrole{o}{\_}~\DUrole{o}{=}~\\
~~~\DUrole{k}{let}~\DUrole{n}{p}~\DUrole{o}{=}~\DUrole{o}{(}\DUrole{n}{fact}~\DUrole{o}{(}\DUrole{n}{inc}~\DUrole{mi}{2}\DUrole{o}{));}~\\
~~~\DUrole{k}{let}~\DUrole{n}{q}~\DUrole{o}{=}~\DUrole{o}{(}\DUrole{n}{fact}~\DUrole{o}{@@}~\DUrole{n}{inc}~\DUrole{mi}{2}\DUrole{o}{);}~\\
~~~\DUrole{k}{let}~\DUrole{n}{r}~\DUrole{o}{=}~\DUrole{o}{(}\DUrole{n}{inc}~\DUrole{mi}{2}~\DUrole{o}{|>}~\DUrole{n}{fact}\DUrole{o}{);}~\\
~~~\DUrole{k}{let}~\DUrole{n}{s}~\DUrole{o}{=}~\DUrole{o}{(}\DUrole{mi}{2}~\DUrole{o}{|>}~\DUrole{n}{inc}~\DUrole{o}{|>}~\DUrole{n}{fact}\DUrole{o}{)}
}
 & 
{\ttfamily \raggedright \noindent
\DUrole{nf}{inc}~\DUrole{ow}{::}~\DUrole{kt}{Int}~\DUrole{ow}{->}~\DUrole{kt}{IO}~\DUrole{kt}{Int}~\\
\DUrole{nf}{fact}~\DUrole{ow}{::}~\DUrole{kt}{Int}~\DUrole{ow}{->}~\DUrole{kt}{IO}~\DUrole{kt}{Int}~\\
~\\
\DUrole{nf}{main}~\DUrole{ow}{=}~\DUrole{kr}{do}~\\
~~~\DUrole{n}{tmp}~\DUrole{ow}{<-}~\DUrole{n}{inc}~\DUrole{mi}{2}\DUrole{p}{;}~\DUrole{n}{p}~\DUrole{ow}{<-}~\DUrole{n}{fact}~\DUrole{n}{tmp}\DUrole{p}{;}~\\
~~~\DUrole{n}{q}~\DUrole{ow}{<-}~\DUrole{p}{(}\DUrole{n}{fact}~\DUrole{o}{=<{}<}~\DUrole{n}{inc}~\DUrole{mi}{2}\DUrole{p}{)}~\\
~~~\DUrole{n}{r}~\DUrole{ow}{<-}~\DUrole{p}{(}\DUrole{n}{inc}~\DUrole{mi}{2}~\DUrole{o}{>{}>=}~\DUrole{n}{fact}\DUrole{p}{)}~\\
~~~\DUrole{n}{s}~\DUrole{ow}{<-}~\DUrole{p}{((}\DUrole{n}{return}~\DUrole{mi}{2}\DUrole{p}{)}~\DUrole{o}{>{}>=}~\DUrole{n}{inc}~\DUrole{o}{>{}>=}~\DUrole{n}{fact}\DUrole{p}{)}
}
 \\
\end{longtable*}

\section{Print debugging%
  \label{print-debugging}%
}

The functional engineer's nightmare: “what actual argument values is
this function really called with at run-time?”

In OCaml, one can readily interleave “print” statements with
the functional code to trace what happens at run-time. In Haskell,
there is an extra issue because “print” statements are only
available in the context of the special recipes with
type \texttt{IO a}, for example constructed
by the do-block syntax, and these do not fit type-wise
in pure functions with regular non-\texttt{IO} types.

There are two ways to achieve this in Haskell. The “pure and magic”
way, and the “impure but obviously effective” way.

The “pure and magic” way is a function called \textbf{trace} in the library
module \texttt{Debug.Trace}. This function has type \texttt{String -> a -> a},
and will print its first argument during evaluation and return its
second argument as result. From the language's perspective, this
function is pure:
\begin{quote}{\ttfamily \raggedright \noindent
\DUrole{kr}{import}~\DUrole{k}{qualified}~\DUrole{nn}{Debug.Trace}~\\
\DUrole{nf}{fact}~\DUrole{n}{n}~\DUrole{ow}{=}~\\
~~~~\DUrole{kr}{let}~\DUrole{n}{n'}~\DUrole{ow}{=}~\DUrole{kt}{Debug}\DUrole{o}{.}\DUrole{kt}{Trace}\DUrole{o}{.}\DUrole{n}{trace}~\DUrole{p}{(}\DUrole{s}{\textquotedbl{}fact:~\textquotedbl{}}~\DUrole{o}{++}~\DUrole{p}{(}\DUrole{n}{show}~\DUrole{n}{n}\DUrole{p}{))}~\DUrole{n}{n}~\DUrole{kr}{in}~\\
~~~~\DUrole{n}{n'}~\DUrole{o}{*}~\DUrole{n}{fact}~\DUrole{p}{(}\DUrole{n}{n'}~\DUrole{o}{-}~\DUrole{mi}{1}\DUrole{p}{)}
}
\end{quote}

In some cases, one may want to do other effectful things beside
printing text.  For example, we may want to print a timestamp or log
output to file.  Besides the other functions from \texttt{Debug.Trace}, you can roll your
own tracing utility using the special function \texttt{unsafePerformIO}:

\setlength{\DUtablewidth}{\linewidth}
\begin{longtable*}[c]{p{0.470\DUtablewidth}p{0.470\DUtablewidth}}

{\ttfamily \raggedright \noindent
\DUrole{k}{let}~\DUrole{n}{f}~\DUrole{n}{args}~\DUrole{o}{...}~\DUrole{o}{=}~\\
~\\
~~~\DUrole{n}{print\_string}~\DUrole{o}{...;}~\\
~~~\DUrole{o}{(...}~\DUrole{k}{value}~\DUrole{n}{computation}~\DUrole{o}{...)}
}
 & 
{\ttfamily \raggedright \noindent
\DUrole{kr}{let}~\DUrole{n}{f}~\DUrole{n}{args}\DUrole{o}{...}~\DUrole{ow}{=}~\\
~~~\DUrole{n}{unsafePerformIO}~\DUrole{o}{\$}~\DUrole{kr}{do}~\\
~~~~~\DUrole{n}{putStr}~\DUrole{o}{...}~\\
~~~~~\DUrole{n}{return}~\DUrole{p}{(}\DUrole{o}{...}~\DUrole{n}{value}~\DUrole{n}{computation}~\DUrole{o}{...}\DUrole{p}{)}
}
 \\
\end{longtable*}

\texttt{unsafePerformIO} takes as single argument a recipe as
constructed eg. by a do-block. When the enclosing expression is
evaluated at run-time, the recipe is first evaluated for effects and
then the final value, given to “\texttt{return}” within the do-block, becomes
the functional value of the enclosing expression. \texttt{unsafePerformIO}
has type \texttt{IO a -> a}.

This is the “impure but obviously effective” way:

\setlength{\DUtablewidth}{\linewidth}
\begin{longtable*}[c]{p{0.470\DUtablewidth}p{0.470\DUtablewidth}}

{\ttfamily \raggedright \noindent
\DUrole{c}{(*~no~import~needed~*)}~\\
~\\
~\\
\DUrole{c}{(*~add~:~int~->~int~->~int~*)}~\\
\DUrole{k}{let}~\DUrole{n}{add}~\DUrole{n}{a}~\DUrole{n}{b}~\DUrole{o}{=}~\\
~\\
~~~~~~\DUrole{n}{print\_endline}~\DUrole{o}{@@}~\DUrole{s2}{\textquotedbl{}add:~\textquotedbl{}}~\\
~~~~~~~~~~~~~~~~~\DUrole{o}{\textasciicircum{}}~\DUrole{o}{(}\DUrole{n}{string\_of\_int}~\DUrole{n}{a}\DUrole{o}{)}~\\
~~~~~~~~~~~~~~~~~\DUrole{o}{\textasciicircum{}}~\DUrole{s2}{\textquotedbl{},~\textquotedbl{}}~\\
~~~~~~~~~~~~~~~~~\DUrole{o}{\textasciicircum{}}~\DUrole{o}{(}\DUrole{n}{string\_of\_int}~\DUrole{n}{b}\DUrole{o}{);}~\\
~~~~~~\DUrole{o}{(}\DUrole{n}{a}~\DUrole{o}{+}~\DUrole{n}{b}\DUrole{o}{)}~\\
~\\
\DUrole{c}{(*~v1~:~int~*)}~\\
\DUrole{k}{let}~\DUrole{n}{v1}~\DUrole{o}{=}~\DUrole{n}{add}~\DUrole{mi}{1}~\DUrole{mi}{1}~\\
~\\
\DUrole{k}{let}~\DUrole{o}{\_}~\DUrole{o}{=}~\\
~~~~~~\DUrole{n}{print\_string}~\DUrole{s2}{\textquotedbl{}hello\textquotedbl{}}\DUrole{o}{;}~\\
~~~~~~\DUrole{n}{print\_int}~\DUrole{n}{v1}\DUrole{o}{;}~\\
~~~~~~\DUrole{n}{print\_int}~\DUrole{o}{@@}~\DUrole{n}{add}~\DUrole{mi}{2}~\DUrole{mi}{3}~\\
~\\
\DUrole{c}{(*~prints~add,~hello,~add~*)}~\\
\DUrole{c}{(*~v1~is~evaluated~where~defined~*)}
}
 & 
{\ttfamily \raggedright \noindent
\DUrole{kr}{import}~\DUrole{nn}{System.IO.Unsafe}~\\
~~~~~~~\DUrole{p}{(}\DUrole{nf}{unsafePerformIO}\DUrole{p}{)}~\\
~\\
\DUrole{nf}{add}~\DUrole{ow}{::}~\DUrole{kt}{Int}~\DUrole{ow}{->}~\DUrole{kt}{Int}~\DUrole{ow}{->}~\DUrole{kt}{Int}~\\
\DUrole{nf}{add}~\DUrole{n}{a}~\DUrole{n}{b}~\DUrole{ow}{=}~\\
~~~~~~\DUrole{n}{unsafePerformIO}~\DUrole{o}{\$}~\DUrole{kr}{do}~\\
~~~~~~~~~~\DUrole{n}{putStrLn}~\DUrole{o}{\$}~\DUrole{s}{\textquotedbl{}add:~\textquotedbl{}}~\\
~~~~~~~~~~~~~~~~\DUrole{o}{++}~\DUrole{p}{(}\DUrole{n}{show}~\DUrole{n}{a}\DUrole{p}{)}~\\
~~~~~~~~~~~~~~~~\DUrole{o}{++}~\DUrole{s}{\textquotedbl{},~\textquotedbl{}}~\\
~~~~~~~~~~~~~~~~\DUrole{o}{++}~\DUrole{p}{(}\DUrole{n}{show}~\DUrole{n}{b}\DUrole{p}{)}~\\
~~~~~~~~~~\DUrole{n}{return}~\DUrole{p}{(}\DUrole{n}{a}~\DUrole{o}{+}~\DUrole{n}{b}\DUrole{p}{)}~\\
~\\
\DUrole{nf}{v1}~\DUrole{ow}{::}~\DUrole{kt}{Int}~\\
\DUrole{nf}{v1}~\DUrole{ow}{=}~\DUrole{n}{add}~\DUrole{mi}{1}~\DUrole{mi}{1}~\\
~\\
\DUrole{nf}{main}~\DUrole{ow}{=}~\DUrole{kr}{do}~\\
~~~~~\DUrole{n}{putStr}~\DUrole{s}{\textquotedbl{}hello\textquotedbl{}}~\\
~~~~~\DUrole{n}{putStr}~\DUrole{o}{.}~\DUrole{n}{show}~\DUrole{o}{\$}~\DUrole{n}{v1}~\\
~~~~~\DUrole{n}{putStr}~\DUrole{o}{.}~\DUrole{n}{show}~\DUrole{o}{\$}~\DUrole{n}{add}~\DUrole{mi}{2}~\DUrole{mi}{3}~\\
~\\
\DUrole{cm}{\{-~prints~hello,~add,~add~-\}}~\\
\DUrole{cm}{\{-~v1's~def~is~not~an~evaluation!~-\}}
}
 \\
\end{longtable*}

Despite its name, this construction is quite safe to use and it
properly witnesses the evaluation of the enclosing expression at
run-time. It is called “unsafe” because it breaks the usual convention
that non-\texttt{IO} functions are pure. However, in the particular case of
print debugging the purity is practically preserved since the I/O
operations do not change the value of the function.

It is bad practice (and strongly frowned upon) to write a Haskell
function using \texttt{unsafePerformIO} that is declared pure via a non-IO
type but whose run-time return value is dependent on run-time side
effects other than its arguments.

(\texttt{unsafePerformIO} is a bit of a taboo. Haskell programmers do not
talk about it to “outsiders”, in the same way that
OCaml programmers do not talk about \texttt{Obj.magic}. But the engineer's
life would be miserable without it. \href{http://www.haskell.org/ghc/docs/latest/html/libraries/base/System-IO-Unsafe.html}{Just be careful}.)

\section{Mutable variables%
  \label{mutable-variables}%
}

OCaml provides the \texttt{ref} type for mutable references to values.
In Haskell, the type \texttt{Data.IORef.IORef} does the same:

\setlength{\DUtablewidth}{\linewidth}
\begin{longtable*}[c]{p{0.470\DUtablewidth}p{0.470\DUtablewidth}}

{\ttfamily \raggedright \noindent
\DUrole{k}{let}~\DUrole{o}{\_}~\DUrole{o}{=}~\\
~~~~\DUrole{k}{let}~\DUrole{n}{v}~\DUrole{o}{=}~\DUrole{n}{ref}~\DUrole{s2}{\textquotedbl{}hello\textquotedbl{}}\DUrole{o}{;}~\\
~~~~\DUrole{k}{let}~\DUrole{n}{s}~\DUrole{o}{=}~\DUrole{o}{!}\DUrole{n}{v}~\DUrole{k}{in}~\DUrole{n}{print\_endline}~\DUrole{n}{s}\DUrole{o}{;}~\\
~~~~\DUrole{n}{v}~\DUrole{o}{:=}~\DUrole{s2}{\textquotedbl{}world\textquotedbl{}}\DUrole{o}{;}~\\
~~~~\DUrole{k}{let}~\DUrole{n}{s}~\DUrole{o}{=}~\DUrole{o}{!}\DUrole{n}{v}~\DUrole{k}{in}~\DUrole{n}{print\_endline}~\DUrole{n}{s}
}
 & 
{\ttfamily \raggedright \noindent
\DUrole{nf}{main}~\DUrole{ow}{=}~\DUrole{kr}{do}~\\
~~~\DUrole{n}{v}~\DUrole{ow}{<-}~\DUrole{n}{newIORef}~\DUrole{s}{\textquotedbl{}hello\textquotedbl{}}~\\
~~~\DUrole{n}{s}~\DUrole{ow}{<-}~\DUrole{n}{readIORef}~\DUrole{n}{v}\DUrole{p}{;}~\DUrole{n}{putStrLn}~\DUrole{n}{s}~\\
~~~\DUrole{n}{writeIORef}~\DUrole{n}{v}~\DUrole{s}{\textquotedbl{}world\textquotedbl{}}~\\
~~~\DUrole{n}{s}~\DUrole{ow}{<-}~\DUrole{n}{readIORef}~\DUrole{n}{v}\DUrole{p}{;}~\DUrole{n}{putStrLn}~\DUrole{n}{s}
}
 \\
\end{longtable*}

The Haskell library \href{http://www.haskell.org/haskellwiki/Library/ArrayRef}{ArrayRef} provides some syntactic sugar to simplify the use of mutable references.

Like in OCaml, the behavior of concurrent access to
mutable variables by different threads in Haskell is not properly
defined by the language. Just avoid data races on \texttt{IORef} and use
Haskell's \texttt{MVar} type to communicate between threads instead.

\section{Arrays%
  \label{arrays}%
}

OCaml's arrays are mutable; although they are less often used, Haskell
supports mutable arrays too. Since in-place array updates are
effectful, in Haskell they must be used in the context of recipes, eg. inside do-blocks:

\setlength{\DUtablewidth}{\linewidth}
\begin{longtable*}[c]{p{0.470\DUtablewidth}p{0.470\DUtablewidth}}

{\ttfamily \raggedright \noindent
\DUrole{k}{let}~\DUrole{o}{\_}~\DUrole{o}{=}~\\
~~~~\DUrole{k}{let}~\DUrole{n}{a}~\DUrole{o}{=}~\DUrole{nn}{Array}\DUrole{p}{.}\DUrole{n}{make}~\DUrole{mi}{10}~\DUrole{mi}{42}\DUrole{o}{;}~\\
~~~~\DUrole{k}{let}~\DUrole{n}{v}~\DUrole{o}{=}~\DUrole{n}{a}\DUrole{o}{.(}\DUrole{mi}{0}\DUrole{o}{);}~\\
~~~~\DUrole{n}{a}\DUrole{o}{.(}\DUrole{mi}{2}\DUrole{o}{)}~\DUrole{o}{<-}~\DUrole{n}{v}~\DUrole{o}{+}~\DUrole{n}{v}\DUrole{o}{;}~\\
~~~~\DUrole{n}{print\_int}~\DUrole{n}{a}\DUrole{o}{.(}\DUrole{mi}{2}\DUrole{o}{)}~\\
~\\
\DUrole{c}{(*~prints~84~*)}
}
 & 
{\ttfamily \raggedright \noindent
\DUrole{nf}{main}~\DUrole{ow}{=}~\DUrole{kr}{do}~\\
~~~~\DUrole{n}{a}~\DUrole{ow}{<-}~\DUrole{n}{newArray}~\DUrole{p}{(}\DUrole{mi}{0}\DUrole{p}{,}\DUrole{mi}{9}\DUrole{p}{)}~\DUrole{mi}{42}~\\
~~~~\DUrole{n}{v}~\DUrole{ow}{<-}~\DUrole{n}{readArray}~\DUrole{n}{a}~\DUrole{mi}{0}~\\
~~~~\DUrole{n}{writeArray}~\DUrole{n}{a}~\DUrole{mi}{2}~\DUrole{p}{(}\DUrole{n}{v}~\DUrole{o}{+}~\DUrole{n}{v}\DUrole{p}{)}~\\
~~~~\DUrole{n}{putStrLn}\DUrole{o}{.}\DUrole{n}{show}~\DUrole{o}{<{}<=}~\DUrole{n}{readArray}~\DUrole{n}{a}~\DUrole{mi}{2}~\\
~\\
\DUrole{cm}{\{-~prints~84~-\}}
}
 \\
\end{longtable*}

The Haskell library \href{http://www.haskell.org/haskellwiki/Library/ArrayRef}{ArrayRef} provides some syntactic sugar to
simplify the manipulation of arrays.

Next to mutable arrays, Haskell provides a lot of different pure array
types in different libraries. Unfortunately, using pure arrays is
quite cumbersome at first. Only later, when you start to grasp that
compositions of functions that operate on arrays are “flattened”
before execution (thanks to non-strict evaluation), does it begin to
make sense. On the plus side, you then can start to write powerful
reusable abstractions with arrays. On the down side, your code then becomes
unreadable.

Decidely, pure Haskell arrays are an acquired taste, one
difficult to share with non-experts.

\section{Type classes in a nutshell%
  \label{type-classes-in-a-nutshell}%
}

A common task in software engineering is to advertise a set of services
using an abstract interface that hides the internal implementation.
For this purpose, OCaml programmers can use objects or parametric modules.

Since Haskell provides neither objects nor parametric modules, Haskell
programmers rely on another mechanism entirely, called “type
classes”. Type classes are nothing like modules but they can help for
encapsulation given the right mindset.

In a nutshell, Haskell type classes express a programming contract
over a set of types (hence the name): that all types in the class,
ie. its “instances”, are guaranteed to provide some other related
functions. Moreover, a class can also provide a default implementation
for some of its functions.

The usual example is the \texttt{Eq} class, written as follows:
\begin{quote}{\ttfamily \raggedright \noindent
\DUrole{kr}{class}~\DUrole{kt}{Eq}~\DUrole{n}{a}~\DUrole{kr}{where}~\\
~~~~\DUrole{p}{(}\DUrole{o}{==}\DUrole{p}{)}~\DUrole{ow}{::}~\DUrole{n}{a}~\DUrole{ow}{->}~\DUrole{n}{a}~\DUrole{ow}{->}~\DUrole{kt}{Bool}~\\
~~~~\DUrole{p}{(}\DUrole{o}{/=}\DUrole{p}{)}~\DUrole{ow}{::}~\DUrole{n}{a}~\DUrole{ow}{->}~\DUrole{n}{a}~\DUrole{ow}{->}~\DUrole{kt}{Bool}~\\
~~~~\DUrole{p}{(}\DUrole{o}{/=}\DUrole{p}{)}~\DUrole{n}{x}~\DUrole{n}{y}~\DUrole{ow}{=}~\DUrole{n}{not}~\DUrole{p}{(}\DUrole{n}{x}~\DUrole{o}{==}~\DUrole{n}{y}\DUrole{p}{)}
}
\end{quote}

This definition expresses the following: “for all types \texttt{a} in the
class \texttt{Eq a}, there exist two operators \texttt{(==)} and \texttt{(/=)} that
accept two arguments of type \texttt{a} and returns a boolean value. Moreover,
the class \texttt{Eq} provides a default implementation of \texttt{(/=)} that uses
the actual implementation of \texttt{(==)} by each particular instance.”

Once a class is specified, a programmer can do two things with it:
\emph{define type instances} of the class, or \emph{define functions over instances}
of the class.

To define an instance of a class, you first define a type, then you define
how the type belongs to the class. For example, one can first define
a new type that implements Peano integers:
\begin{quote}{\ttfamily \raggedright \noindent
\DUrole{kr}{data}~\DUrole{kt}{Peano}~\DUrole{ow}{=}~\DUrole{kt}{Zero}~\DUrole{o}{|}~\DUrole{kt}{Succ}~\DUrole{kt}{Peano}
}
\end{quote}

Then, express that \texttt{Peano} is a member of \texttt{Eq}:
\begin{quote}{\ttfamily \raggedright \noindent
\DUrole{kr}{instance}~\DUrole{kt}{Eq}~\DUrole{kt}{Peano}~\DUrole{kr}{where}~\\
~~~~\DUrole{p}{(}\DUrole{o}{==}\DUrole{p}{)}~\DUrole{kt}{Zero}~\DUrole{kt}{Zero}~\DUrole{ow}{=}~\DUrole{kt}{True}~\\
~~~~\DUrole{p}{(}\DUrole{o}{==}\DUrole{p}{)}~\DUrole{p}{(}\DUrole{kt}{Succ}~\DUrole{n}{x}\DUrole{p}{)}~\DUrole{p}{(}\DUrole{kt}{Succ}~\DUrole{n}{x}\DUrole{p}{)}~\DUrole{ow}{=}~\DUrole{p}{(}\DUrole{n}{x}~\DUrole{o}{==}~\DUrole{n}{x}\DUrole{p}{)}~\\
~~~~\DUrole{p}{(}\DUrole{o}{==}\DUrole{p}{)}~\DUrole{kr}{\_}~\DUrole{kr}{\_}~\DUrole{ow}{=}~\DUrole{kt}{False}
}
\end{quote}

Once this definition is visible, automatically any expression of the
form “\texttt{x /= y}” becomes valid, thanks to the default implementation
of \texttt{(/=)} already provided by \texttt{Eq}.

Given a type class, one can also define functions over any possible
instance of that class. For example, given \texttt{Eq}, one can write a function
which for any three values, returns \texttt{True} if at least two are equal:
\begin{quote}{\ttfamily \raggedright \noindent
\DUrole{nf}{any2}~\DUrole{ow}{::}~\DUrole{p}{(}\DUrole{kt}{Eq}~\DUrole{n}{a}\DUrole{p}{)}~\DUrole{ow}{=>}~\DUrole{n}{a}~\DUrole{ow}{->}~\DUrole{n}{a}~\DUrole{ow}{->}~\DUrole{n}{a}~\DUrole{ow}{->}~\DUrole{kt}{Bool}~\\
\DUrole{nf}{any2}~\DUrole{n}{x}~\DUrole{n}{y}~\DUrole{n}{z}~\DUrole{ow}{=}~\DUrole{p}{(}\DUrole{n}{x}~\DUrole{o}{==}~\DUrole{n}{y}\DUrole{p}{)}~\DUrole{o}{||}~\DUrole{p}{(}\DUrole{n}{y}~\DUrole{o}{==}~\DUrole{n}{z}\DUrole{p}{)}~\DUrole{o}{||}~\DUrole{p}{(}\DUrole{n}{x}~\DUrole{o}{==}~\DUrole{n}{z}\DUrole{p}{)}
}
\end{quote}

Notice the prefix “\texttt{(Eq a) =>}” in the type signature. This means
that “the function \texttt{any2} is defined over any type \texttt{a} as long as
\texttt{a} is an instance of \texttt{Eq}”.

Type classes can inherit all the operations of another class, before
it defines its owns. For example, the class \texttt{Ord} defines operator
\texttt{(<=)} (less than or equal), but for this it requires \texttt{(==)}
defined by \texttt{Eq}. This is written as follows:
\begin{quote}{\ttfamily \raggedright \noindent
\DUrole{kr}{class}~~\DUrole{p}{(}\DUrole{kt}{Eq}~\DUrole{n}{a}\DUrole{p}{)}~\DUrole{ow}{=>}~\DUrole{kt}{Ord}~\DUrole{n}{a}~~\DUrole{kr}{where}~\\
~~~~\DUrole{p}{(}\DUrole{o}{<=}\DUrole{p}{)}~~\DUrole{ow}{::}~\DUrole{n}{a}~\DUrole{ow}{->}~\DUrole{n}{a}~\DUrole{ow}{->}~\DUrole{kt}{Bool}~\\
~~~~\DUrole{cm}{\{-~some~other~Ord~definitions~omitted~here~-\}}
}
\end{quote}

Type classes can be used directly to implement type-safe “container” data structures
over arbitrary other types. For example, an abstract interface for an associative
array over arbitrary keys can be defined with:
\begin{quote}{\ttfamily \raggedright \noindent
\DUrole{kr}{class}~\DUrole{p}{(}\DUrole{kt}{Ord}~\DUrole{n}{k}\DUrole{p}{)}~\DUrole{ow}{=>}~\DUrole{kt}{Map}~\DUrole{n}{mt}~\DUrole{n}{k}~\DUrole{n}{v}~\DUrole{kr}{where}~\\
~~~~\DUrole{n}{empty}~\DUrole{ow}{::}~\DUrole{n}{mt}~\DUrole{n}{k}~\DUrole{n}{v}~\\
~~~~\DUrole{n}{insert}~\DUrole{ow}{::}~\DUrole{n}{k}~\DUrole{ow}{->}~\DUrole{n}{v}~\DUrole{ow}{->}~\DUrole{n}{mt}~\DUrole{n}{k}~\DUrole{n}{v}~\DUrole{ow}{->}~\DUrole{n}{mt}~\DUrole{n}{k}~\DUrole{n}{v}~\\
~~~~\DUrole{n}{delete}~\DUrole{ow}{::}~\DUrole{n}{k}~\DUrole{ow}{->}~\DUrole{n}{mt}~\DUrole{n}{k}~\DUrole{n}{v}~\DUrole{ow}{->}~\DUrole{n}{mt}~\DUrole{n}{k}~\DUrole{n}{v}~\\
~~~~\DUrole{p}{(}\DUrole{o}{!}\DUrole{p}{)}~\DUrole{ow}{::}~\DUrole{n}{mt}~\DUrole{n}{k}~\DUrole{n}{v}~\DUrole{ow}{->}~\DUrole{n}{k}~\DUrole{ow}{->}~\DUrole{n}{v}
}
\end{quote}

This definition creates a new type class called “\texttt{Map}”, whose type
instances are all of the form “\texttt{mt k v}” (this means that all
instances of \texttt{Map} must be parametric with two type parameters). It
also requires that the 1st type parameter (\texttt{k}, the key type) of its
instances be a member of class \texttt{Ord}. For each instance \texttt{mt k v},
\texttt{Map} provides four services \texttt{empty}, \texttt{insert}, \texttt{delete} and
\texttt{(!)}.

Once this definition is visible, it becomes possible to implement one
or more concrete associative array data structures, and make them
instances of \texttt{Map} via suitable \texttt{instance} declarations. From the
perspective of third-party modules, the only visible services of these
concrete implementations are those provided by the class.

\section{Rolling your own recipe system%
  \label{rolling-your-own-recipe-system}%
}

The previous sections have emphasized how “lists of statements” or
“recipes” must be defined as constant data structures by the program,
and are later consumed by the run-time system during execution,
starting from \texttt{main}, to produce effects.

Once you start to become comfortable with composing these recipes
using do-blocks, the \texttt{<-} binding and the \texttt{(>{}>=)} and
\texttt{(=<{}<)} operators, a question often arises: \emph{is it possible to write one own's
recipe type and interpreter?}

The answer is yes! To do so, you will need to define:
\begin{itemize}

\item a parametric type \texttt{M a} that represents a “recipe of statements that eventually
produce a value of type \texttt{a}”;

\item a function \texttt{return :: a -> M a} that takes as argument a value, and constructs
a recipe intended to produce that value later;

\item a function \texttt{(>{}>=) :: M a -> (a -> M b) -> M b} that takes as
arguments one recipe and a constructor for another recipe, and
“connects” the two recipes together;

\item optionally, one or more effectful statements of type \texttt{M a} that
can be used in recipes next to \texttt{return};

\item optionally, an interpreter for objects of type \texttt{M a}, that performs
its effects in the way you see fit.

\end{itemize}

Once you have defined the first three properly (see below), the
do-block syntax presented earlier is automatically extended to your
new type, inferring the right types from your custom definition of
\texttt{return}.

Here are some useful recipe systems already available in Haskell:
\begin{itemize}

\item \texttt{IO a}: “recipes” evaluated by the Haskell run-time
system for global effects during execution. The peculiarity of this
one is that its implementation is completely hidden from the
language; you can't redefine \texttt{IO} directly in Haskell. The
effectful statements of this recipe system include \texttt{putStr} and
\texttt{getChar}, already presented above.

\item \texttt{Maybe a}: “recipes” that eventually provide a value of type
\texttt{a}, but where the evaluation stops automatically at the first
intermediate statement that returns \texttt{Nothing}. The effectful
statements of this recipe system include \texttt{fail}, which forces
\texttt{Nothing} to be generated during the effectful evaluation.

\item \texttt{Writer l a}: “recipes” that eventually provide a value of type
\texttt{a}, but where the evaluation keeps a log of messages generated by
the statements. The effectful statements of this recipe system
include \texttt{tell}, which generates a message for the log during
effectful evaluation.

\item \texttt{State s a}: “recipes” that eventually provides a value of type
\texttt{a}, but where each intermediate statement can modify an internal
value of type \texttt{s} (a state). The effectful statements of this
recipe system include \texttt{get} and \texttt{set}, which can access and
modify the internal state during effectful evaluation.

\end{itemize}

\emph{About the M-word}: There is a word used in the Haskell community
which starts with the letter “M” and causes a great deal of confusion
and misunderstandings.  I wish to avoid using the M-word entirely in
this document. I believe that using and understanding the M-word is
unnecessary to learn how to write Haskell programs productively.

Nevetherless, you should understand that whenever you read something
about the M-word, it really refers to what I explained in this
section.  When you read “the type \texttt{T a} is a M...”, it
really means that “the type \texttt{T a} describes recipes of statements that
produce values of type \texttt{a} during evaluation” and also that “the type \texttt{T a} is
an instance of the type class \texttt{M...}, which provides the services \texttt{return}
and \texttt{(>{}>=)}”.

Likewise, when you read or hear “let's define a M...”, this simply
refers to the act of writing a definition for a new custom recipe type and
two new functions \texttt{return} and \texttt{(>{}>=)}, and then using
\texttt{instance} to declare an instance of the \texttt{M} class with them.

The reason why Haskell programmers \emph{eventually} end up caring a great
deal about the M-word, in the same way they end up caring about the
\texttt{Applicative}, \texttt{Functor} and \texttt{Monoid} classes, is that there are
very good software components that can be defined using only the
services of these classes. This means these software components are
greatly reusable, because they apply automatically to all types later
declared to be instances of these classes.

\section{Haskell limitations%
  \label{haskell-limitations}%
}

Of course, there is also a price to pay. Haskell was not primarily
designed to be a functional language by engineers for engineers. There
are three features that OCaml gets “just right” and are unfortunately
completely missing in Haskell:
\begin{itemize}

\item parametric modules and module composition;

\item named and optional function parameters;

\item polymorphic variant types.

\end{itemize}

Haskell practitioners eventually develop some other conventions
(“best practices”) to achieve parametric modularity without parametric
modules. This usually involves mixing and matching the following features:
\begin{itemize}

\item preprocessing using the C preprocessor (\texttt{ghc -cpp}), which can be used
to mimic parametric (but not composable) modules in the same way as in C;

\item record types, placing the module's type definition in ghost type parameters
of a record type, and the module's methods in its fields;

\item type classes, eg. to replace OCaml's \texttt{Map} functor and
\texttt{OrderedType} parameter, instantiated as \texttt{IntMap}, \texttt{BoolMap},
etc., by \texttt{Map} and \texttt{Ord} type classes (as suggested above).

\end{itemize}

Named and optional function parameters, in practice, become less
important once the Haskell practioner knows how to fully leverage
higher-order functions and non-strict evaluation. A programming
technique to implement \href{http://neilmitchell.blogspot.nl/2008/04/optional-parameters-in-haskell.html}{optional function arguments using record
types} and default record fields was described by Neil Mitchell
in 2008.

Polymorphic variant types (\texttt{{[}> `A | `B{]}}) are, unfortunately, without direct equivalent in
Haskell. Haskell is a “closed world language” where type inference can
only succeed once all types from all compilation units are known. However, a language
extension from the GHC compiler called \href{http://www.haskell.org/haskellwiki/GHC/Type_families}{Type families} can be abused to provide
somewhat similar functionality as OCaml's polymorphic variants.

\section{External links%
  \label{external-links}%
}

\subsection{Searching for more information%
  \label{searching-for-more-information}%
}
\begin{itemize}

\item The \href{http://www.haskell.org/haskellwiki/}{Haskell wiki}.

\item The \href{https://en.wikibooks.org/wiki/Haskell}{Haskell Wikibook}.

\item \href{http://www.haskell.org/hoogle/}{Hoogle}: an API search engine, able to search by function name or by type signature.

\item The \texttt{\#haskell} IRC channel on FreeNode.

\item The \href{http://www.haskell.org/ghc/docs/latest/html/users_guide/}{GHC Manual}.

\end{itemize}

\subsection{Further reads%
  \label{further-reads}%
}
\begin{itemize}

\item Paul Hudak, John Peterson, Joseph Fasel, and Reuben Thomas. \href{http://www.haskell.org/tutorial/}{A
Gentle Introduction to Haskell, Version 98}. June 2000. Also known
as the “official Haskell tutorial”.

\item Miran Lipovaca. \href{http://learnyouahaskell.com/}{Learn You a Haskell for Great Good!}. April 2011. Also known
as “the funkiest way to learn Haskell”, but a very pleasant read.

\item Bryan O'Sullivan, Don Stewart, John Goerzen. \href{http://book.realworldhaskell.org/}{Real World
Haskell}. November 2008. With lots of practical examples.

\item Edward Z. Yang, \href{http://blog.ezyang.com/2013/04/resource-limits-for-haskell/}{Resource limits for
Haskell}. April 2013. Explains how to monitor and control space usage
in Haskell functions, including during non-strict evaluation on
parallel computers.

\item Robert Harper, \href{http://existentialtype.wordpress.com/2011/04/09/persistence-of-memory/}{Persistence of Memory}, April 2011. Promotes the
use of persistent (immutable) data structures, and argues that the
common debate about run-time efficiency of mutable vs. immutable
data structures is often misdirected.

\item Robert Harper, \href{http://existentialtype.wordpress.com/2011/04/24/the-real-point-of-laziness/}{The Point of Lazyness}, April 2011. Argues that
non-strict evaluation is desirable even in eager languages, at least
to manage processes and streams elegantly.

\item Lennart Augustsson, \href{http://augustss.blogspot.nl/2011/05/more-points-for-lazy-evaluation-in.html}{More points for lazy evaluation},
May 2011. Argues that non-strict evaluation promotes reuse of
software components.

\item Edward Z. Yang, \href{http://blog.ezyang.com/2014/01/so-you-want-to-add-a-new-concurrency-primitive-to-ghc/}{So you want to add a new concurrency primitive to
GHC...}, January 2014. Explains how GHC's Haskell still has
conceptual issues internally with the notion of mutable memory.

\item Andreas Voellmy, Junchang Wang, Paul Hudak and Kazuhiko
Yamamoto. \href{http://haskell.cs.yale.edu/wp-content/uploads/2013/08/hask035-voellmy.pdf}{Mio: A high-performance multicore IO manager for
GHC}. September 2013. Explains how GHC implements \texttt{IO}
internally, and how it was recently extended to better support
parallel execution.

\item Ben Rudiak-Gould, Alan Mycroft and Simon Peyton Jones. \href{https://research.microsoft.com/en-us/um/people/simonpj/papers/not-not-ml/index.htm}{Haskell is
Not Not ML}. 2006. Suggests that there is an underlying common
language behind Haskell and SML, that can run programs written in
either. Also \href{http://lambda-the-ultimate.org/node/1248}{discussed by Ehud Lamm here}.

\end{itemize}

\subsection{References%
  \label{references}%
}
\begin{itemize}

\item Edward Z. Yang, \href{http://blog.ezyang.com/2010/10/ocaml-for-haskellers/}{OCaml for Haskellers}, October 2010.

\item Xavier Leroy et al., \href{http://caml.inria.fr/pub/docs/manual-ocaml-4.01/}{The OCaml system release 4.01}, September 2013.

\end{itemize}

\DUtransition

\section{Copyright and licensing%
  \label{copyright-and-licensing}%
}

Copyright © 2014, Raphael ‘kena’ Poss.
Permission is granted to distribute, reuse and modify this document
according to the terms of the Creative Commons Attribution-ShareAlike
4.0 International License.  To view a copy of this license, visit
\url{http://creativecommons.org/licenses/by-sa/4.0/}.

\DUtransition

\href{http://www.structured-commons.org}{SC} fingerprint: \texttt{fp:QU8zQlDz72n5SSapjIGsHw39Fu7IfyrARcc549DmTzY7CA}

\end{document}